\documentclass[usenatbib]{mn2e}

\usepackage{amsmath}

\usepackage{graphicx}
\usepackage{txfonts}
\usepackage{bm}
\usepackage{natbib}

\usepackage{hyperref}
\hypersetup{colorlinks=true,
            citecolor=blue,
            linkcolor=blue}

\def\be {\begin{equation}}
\def\ee {\end{equation}}

\def\tt {\{t\}}

\def\nul#1{}

\begin{document}
%
\title[On the Jacobi capture origin of binaries]{On the Jacobi capture origin of binaries with applications to the Earth-Moon system and black holes in galactic nuclei}

\author[Boekholt, Rowan and Kocsis]{Tjarda C. N. Boekholt$^{1}$\thanks{E-mail: tjarda.boekholt@physics.ox.ac.uk}, Connar Rowan$^{1}$\thanks{E-mail: connar.rowan@physics.ox.ac.uk} and Bence Kocsis$^{1}$\thanks{E-mail: bence.kocsis@physics.ox.ac.uk}\\
$^{1}$Rudolf Peierls Centre for Theoretical Physics, Clarendon Laboratory, Parks Road, Oxford, OX1 3PU, UK\\
}

\date{\today}

\maketitle

\begin{abstract}
Close encounters between two bodies in a disc often result in a single orbital deflection. However, within their Jacobi volumes, where the gravitational forces between the two bodies and the central body become competitive, temporary captures with multiple close encounters become possible outcomes: a Jacobi capture. We perform 3-body simulations in order to characterise the dynamics of Jacobi captures in the plane. We find that the phase space structure resembles a Cantor-like set with a fractal dimension of about 0.4. The lifetime distribution decreases exponentially, while the distribution of the closest separation follows a power law with index 0.5. In our first application, we consider the Jacobi capture of the Moon. We demonstrate that both tidal captures and giant impacts are possible outcomes. The impact speed is well approximated by a parabolic encounter, while the impact angles follow that of a uniform beam on a circular target. Jacobi captures at larger heliocentric distances are more likely to result in tidal captures. In our second application, we find that Jacobi captures with gravitational wave dissipation can result in the formation of binary black holes in galactic nuclei. The eccentricity distribution is approximately super-thermal and includes both prograde and retrograde orientations. 
We conclude that dissipative Jacobi captures form an efficient channel for binary formation, which motivates further research into establishing the universality of Jacobi captures across multiple astrophysical scales.
\end{abstract}



\section{Introduction}

\subsection{Binary formation channels}

There is a rich variety of formation channels for binary systems. The Earth-Moon binary is hypothesised to have formed after one or more major impacts \citep[e.g.][]{Hartmann75, Benz86, Canup_2008,2012Sci...338.1047C, 2015Natur.520..212M, 2017NatGe..10...89R,Cano_2020}. 
Some of the moons in the Solar System with eccentric and inclined orbits are likely to have been captured dynamically, such as Saturn's moon Phoebe \citep{Johnson05} and Neptune's moon Triton \citep{Agnor06, Nesvorny07}. Binary stars can form in-situ in a fragmenting natal gas cloud \citep[e.g.][]{Bate95, 2018MNRAS.479..667R}, or from the fission of optically thick protostars \citep{Lucy77}. Unstable triple systems reduce to a binary when one of the objects is dynamically ejected \citep[e.g.][]{Heggie75, Bahcall83, 2022ApJ...925..178H, 2021arXiv210804272T}. A pathway for forming exotic binaries is through the effects of stellar evolution in hierarchical triples, leading up to mergers through mass transfer or dynamical instability \citep{TEDI2012}, finally leaving a binary consisting of one original component and a rejuvenated one, e.g. a blue straggler \citep[e.g.][]{Perets09, Naoz14, spz16}. New binaries can also form from old ones through an exchange \citep{1976MNRAS.175P...1H}. This occurs when a stellar binary is tidally disrupted by a supermassive black hole, where one star is ejected at runaway speeds, while the remaining star is now bound to the black hole \citep{Hills88}. This scenario also applies to binary asteroids encountering planets \citep[e.g.][]{philpott10}. 

Forming a binary system out of two single bodies on a hyperbolic trajectory requires some form of dissipation of orbital energy during pericenter passage. For example, interstellar comets are captured by the Sun if sufficient orbital energy can be extracted by Jupiter \citep[e.g.][]{1982ApJ...255..307V, 2021PSJ.....2...53N}. For other systems where there are no mediating bodies present, dissipation can occur by other means, such as tidal deformation. For this mechanism to be efficient however, the two interacting bodies would need to approach each other up to a few times their own physical radii \citep[e.g.][]{fabian75, press77, mardling01}. Close approaches are also required for successful gravitational wave captures of black holes and other stellar remnants \citep[e.g.][]{hansen72, 2021PhRvD.104h3020B, 2021arXiv210605575G}. Alternatively, if the bodies are embedded in some gaseous medium, gas drag or friction \citep[e.g.][]{1999ApJ...513..252O} can potentially dissipate the orbital energy \citep[e.g.][]{2005ApJ...630..152E}. For example, gas dynamical friction in proto-planetary discs can drive the formation of binary planetesimals, and even their subsequent merger \citep{Grishin2016}. On top of that, gas mass accretion onto the bodies also aids in increasing their binding energy. Irrespective of the dissipation mechanism, binary formation out of two single bodies would potentially become much more efficient if they had the opportunity of experiencing multiple close encounters, increasing the probability of eventually dissipating sufficient orbital energy. 
However, this is only possible if the two bodies are able to exchange energy with a background potential, such as a Keplerian disc. 

\subsection{The Jacobi capture channel}

Encounters in Keplerian discs have been extensively studied within the context of (temporary) satellite capture \citep[e.g.][]{Heppen77, Nishida83, Petit86,  Murison89, Astakhov04, Suetsugu11, Higuchi_2016, Higuchi_2017, Qi2018}.
If the satellite is treated as a mass-less body, the dynamics can be studied with the restricted three-body problem \citep[e.g.][]{colombo66, Araujo2008, Assis2014}. The conserved quantity in this case is the Jacobi integral and zero-velocity contours in the co-rotating frame 
distinguish a region around the planet in which the satellite is more strongly bound to the planet than to the star \citep[e.g.][Fig.~7-8]{binney87}. The radius of this Jacobi ellipsoid, also known as the Jacobi radius or Hill radius, is approximately given by

\begin{equation}
    \label{eq:R_J}
    R_H = r \left( \frac{m}{3M} \right)^{\frac{1}{3}},
\end{equation}

\noindent with $r$ the orbital radius, $M$ the mass of the central body and $m$ the mass of the secondary. A Jacobi capture \citep{Singer68} occurs when the third object is captured within the Jacobi volume of the secondary for a prolonged duration. As discussed by \cite{Singer68, Singer86}, a variety of trajectories exist with a distribution of Jacobi capture lifetimes, potentially up to infinite time. 

Hill's problem for circular and planar orbits is the simplest case, where typically the offset in semi-major axis of the two interacting bodies, i.e. the impact parameter \citep{Petit86}, is varied systematically. Galleries of orbits with varying impact parameters are presented by e.g. \citet[][]{Nishida83, Petit86, Iwasaki07}. Rather than plotting the capture orbits, one can also plot statistical properties, such as capture lifetime and closest encounter, as a function of impact parameter. These curves exhibit an interesting structure consisting of various baselines, islands and superposed peaks and dips, somewhat resembling emission/absorption line spectra. Examples of such ``dynamical spectra'' can be found in \citet[][Fig. 4]{Petit86}, \citet[][Fig. 16]{Ida89} and \citet[][Fig. 8]{Iwasaki07}. 
High resolution sampling of the impact parameter space reveals the presence of ``transitional islands'' \citep{Petit86, Iwasaki07}. The origin of these transitional regions is thought to be the crossing of asymptotically periodic orbits \citep{Petit86}, i.e. infinitely long captures.  Such a crossing causes the system to enter a new orbit family, resulting in a discontinuous change in the spectra \citep[e.g.][Fig.~8]{Petit86}. Outgoing orbits might cross multiple unstable periodic orbits, resulting in a hierarchical structure of transitions of
higher and higher order \citep{Petit86}. This suggests that the true transitions have a self-similar and Cantor-like structure \citep{Petit86}. 
The chaotic nature of Jacobi captures was also revealed by \cite{Murison89}, who performed numerical simulations of the phenomenon, and produced Poincar\'e surfaces of section. These showed both periodic orbit families, i.e. ``islands'' in phase space, and chaotic regions. Self-similarity was demonstrated as well, indicating the fractal structure of the phase space associated with capture. 
The presence of chaotic layers which delineate single flyby orbits from periodic ones, is thought to be at the heart of the trapping process of satellites within the Hill sphere \citep{Astakhov04}.
Furthermore, the chaotic layer also facilitates the reorientation of orbits between prograde and retrograde, as well as producing relatively high inclinations \citep{Astakhov04}. 
\citet{Heppen77} studied another aspect of the chaotic orbital evolution during capture, in the context of libration capture of Jupiter's satellites. They reveal two different orbital modes, where in the first one, the orbit keeps approaching the zero-velocity orbit closely, thereby maintaining a probability to escape. In the second mode, the orbit has changed such that it remains at a distance from the zero-velocity orbit. The presence of the second mode can drastically increase the capture lifetime by an order of magnitude or more. The chaotic nature and extended duration of Jacobi captures, in combination with some form of orbital dissipation, forms a potentially efficient channel for binary formation with applications to various astrophysical systems.

Jacobi captures with non-zero eccentricities and inclinations have been investigated amongst others by \citet{Suetsugu11}.
They define four different types of capture orbits depending on the pre-capture heliocentric orbital eccentricity and energy integral.
In the regime of small eccentricities (which we will focus on in this paper), they find that
the capture orbit is typically retrograde in the vicinity of the planet's Hill sphere. 
These retrograde bodies mostly originate from orbits near the planetary orbit, while prograde bodies are captured from further out \citep{Higuchi_2016}. 
The latter study also detects an asymmetry in the sense that bodies which are initially closer to the central star are more easily captured than bodies at larger orbital radii \citep{Higuchi_2016}. 
Also in the regime of near-zero relative velocities, it has been shown that mass ratio (in the restricted problem) has no strong effect on the Jacobi capture trajectories \citep{Heppen77}.

Another application of Jacobi captures in the Solar System is wide binaries in the Kuiper belt \citep{Goldreich_2002, 2005MNRAS.360..401A, 2007MNRAS.379..229L, Schlichting_2008a, Schlichting_2008b}. Two scenarios are provided for making the temporary capture a permanent one. In the L3 scenario, a third body enters the interaction, taking away the excess orbital energy. Alternatively, in the  L2s scenario the two bodies are embedded in a sea of planetesimals, so that dynamical friction becomes the effective dissipation mechanism. Another example is the famous impact of the object Shoemaker-Levy 9 (SL-9) onto Jupiter \citep{Benner_1995}. Due to its short Lyapunov time scale of about 10 years, its past orbit can only be determined statistically. This revealed that SL-9 spent several decades, or even longer, within Jupiter's Jacobi volume, before a close encounter resulted in the tidal disruption.

Planets embedded in proto-planetary discs might also be susceptible to Jacobi captures. Depending on the specific interaction between the gas and the planets, various migration types can be defined \citep[e.g.][]{1986ApJ...309..846L, 2010exop.book..347L}. As a result, planets can move inwards or outwards \citep[e.g.][]{2000MNRAS.315..823P, 2002ApJ...565.1257T}, rendering close planetary encounters inevitable.
Contrary to the restricted case, the two encountering bodies are of similar mass, and it is often assumed that the outcome of the interaction is an ejection of one of the bodies (becoming unbound and free-floating, or bound and very eccentric), while the other body potentially becomes a short-period planet \citep[e.g.][]{2008ApJ...686..580C}. Encounters of this type are hypothesised to have occurred between Mercury and Venus \citep{2020MNRAS.496.3781F}. Alternatively, Jacobi captures can lead to  temporary binary configurations. When tidal and/or gas dissipation is taken into account during the interaction, this potentially leads to the formation of binary planets, the capture of (large) moons, or an enhanced rate of (giant) impacts \citep[e.g.][]{nakazawa83}.
Apart from orbital dissipation mechanisms, permanent capture can also result from mass-change effects \citep{Heppen77}.
For example, mass loss from the Sun and/or mass accretion by Jupiter increases the stability of a captured satellite.

Keplerian encounters as described above are also thought to occur in galactic nuclei. Stars and compact remnants orbit the central supermassive black hole and are prone to close encounters. Binary-single interactions in a Keplerian potential provide a plausible explanation for several S-stars \citep{Trani19a}, but also play an important role in triggering the
coalescence of compact binaries around supermassive black holes \citep{Trani19b}. The source of dissipation in these cases comes from three-body dynamical interactions and/or gravitational wave emission. A potentially very fruitful Jacobi capture environment is provided by accretion discs in active galactic nuclei (AGN). Interactions between black holes with added gas effects is thought to catalyse the formation of gravitational wave sources \citep{Barry2012}. In this case, the two encountering bodies are stellar mass black holes or other stellar remnants, which orbit the supermassive black hole. The stellar mass objects are thought to form in-situ within the accretion disc from evolved massive stars, or to have been captured after repeated disc crossings \citep{Bartos17,2017MNRAS.464..946S}. The latter occurs naturally 
due to gas dynamical friction and 
vector resonant relaxation, which tends to flatten the black hole distribution within the nuclear star cluster \citep[e.g.][]{2018MNRAS.476.4224P, 2018PhRvL.121j1101S,2021ApJ...919..140S,2021arXiv211109011M}. The population of single black holes within the AGN disc is also prone to migration \citep[e.g.][]{2011PhRvD..84b4032K,2011MNRAS.417L.103M}, thus naturally leading to Keplerian encounters, similar to the planetary case described above.   
Numerous single-single encounters with the added ingredients of energy dissipation due to the gas or gravitational waves \citep[e.g.][]{1987ApJ...321..199Q,OLeary_2009,2020ApJ...898...25T,2021MNRAS.tmp.1661G, Li2022}, 
are thought to naturally lead to a significant population of embedded binary black holes. Subsequent evolution through binary-single encounters and gas dynamical friction further hardens the binary \citep[e.g.][]{2017MNRAS.464..946S,2020arXiv201009765S}, until the eventual gravitational wave in-spiral and merger event detectable by LIGO, Virgo and KAGRA \citep{2016PhRvX...6d1015A, 2018LRR....21....3A}, as well as the upcoming LISA \citep{2017arXiv170200786A}. The AGN channel naturally produces black hole mergers with a non-zero eccentricity ($> 0.1$ for the most massive black holes \citep{2018ApJ...860....5G, 2020arXiv201009765S,2021ApJ...907L..20T}). Furthermore, multiple generation mergers lead to the formation of black holes in the upper mass gap, which cannot be explained by single stellar evolution \citep{2020ApJ...903..133S,2021ApJ...908..194T}. 

The treatment of the capture process of two single black holes into a binary is often (too) simplified. It is commonly assumed that if the objects approach each other within a distance of order the Hill radius, they will dissipate a sufficient amount of energy to become bound. On top of that, a single flyby should already provide sufficient opportunity for binaries to form. As described previously in this section, the capture process in Keplerian discs is highly chaotic with an underlying fractal structure of the phase space. The orbital motions are thus expected to be complex and a capture is not guaranteed to be the final outcome. This motivates a detailed, numerical study on the contribution of Jacobi captures to the formation of binary black holes in AGN.

\subsection{Outline and novelty}

This paper is the first in a series whose aim is to determine the universality of dissipative Jacobi captures and its manifestations in a wide variety of astrophysical systems. Here, we provide a benchmark study where we exclude dissipation in the simulations, and where we constrain the system to consist of point-particles confined to the plane on initially circular orbits. In later papers, we will gradually build up the complexity of the model by adding a three-dimensional velocity dispersion, dissipative terms in the numerical integration, and other physical ingredients in order to address specific systems. 

The experimental setup in this work resembles that of e.g. \citet{Iwasaki07}, where statistical properties of Jacobi captures, such as closest approach and lifetime, are measured as a function of impact parameter. Our aim is to build upon these results and to improve them by using arbitrary-precision N-body calculations \citep{2015ComAC...2....2B,2021PhRvD.104h3020B}. By reducing the numerical noise in the dynamical spectra, we will reveal a clear self-similar structure, which provides further empirical evidence for the underlying Cantor set structure mentioned by \citet{Petit86}. We also provide a novel intuitive picture for the transitional islands in the Jacobi capture phase space, by correlating them with three orbit families (see Figs.~\ref{fig:spectra_ne}, \ref{fig:spectra_dr} and \ref{fig:orbit_types}, and Sec.~\ref{sec:physical_picture}). Using our new high-resolution spectra, we measure one-dimensional cross sections, i.e. line sections, for various statistical quantities of interest, such as number of close approaches, closest separation and orientation (prograde vs. retrograde). We apply these results not only to a planetary system with tides (Sun-Earth-Moon), but also to black holes in galactic nuclei in the presence of gravitational wave emission. The latter experiment demonstrates that Jacobi captures can act as a formation channel for gravitational wave sources \citep[see also][]{Li2022}.

We describe the experimental setup and numerical tools in Sec.~\ref{sec:methods}.  We present our results on the conservative dynamics of Jacobi captures in Sec.~\ref{sec:result}. The effects of dissipation are post-processed in Sec.~\ref{sec:dissipative}, with applications to the Sun-Earth-Moon system and gravitational wave captures of black holes in AGN. We finish by discussing the caveats of the model and our conclusions in Sec.~\ref{sec:caveats}. 

\section{Methods}\label{sec:methods}

\begin{figure}
\centering
\begin{tabular}{c}
\includegraphics[height=0.374\textwidth,width=0.5\textwidth]{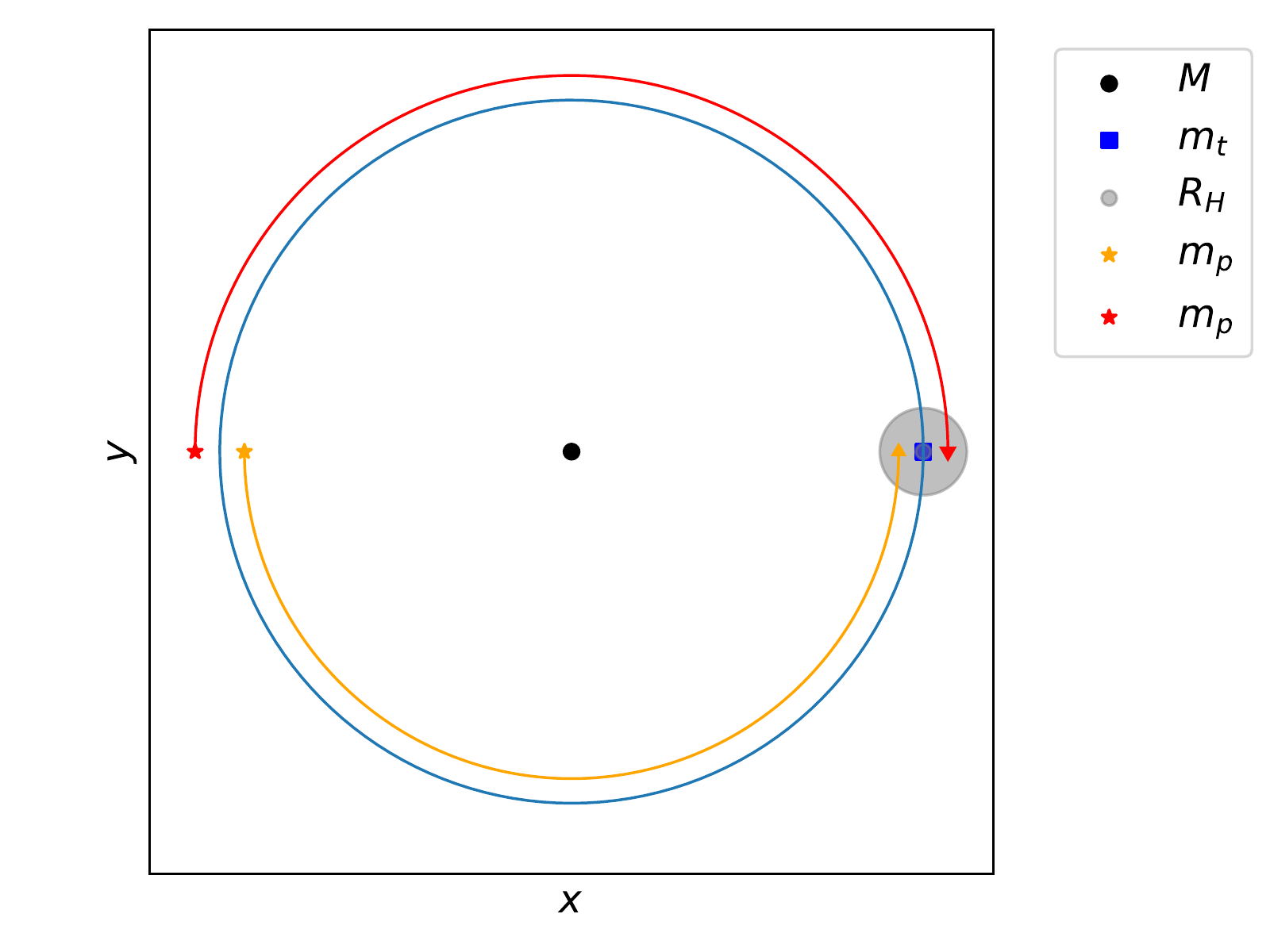} \\
\end{tabular}  
\caption{ Schematic illustration of the experimental setup. The frame is centralised on the central body ($M$), and co-rotating with the target body ($m_t$). The target's Hill radius ($R_H$) is shaded in grey. For the projectile body (with mass $m_p$), we show two orbits with positive (red star) and negative (orange star) impact parameters, which both cross the target's Hill radius.  }
\label{fig:exp_setup}
\end{figure}

A schematic illustration of our experimental setup is presented in Fig.~\ref{fig:exp_setup}. The configuration consists of a central body with mass $M$, which is orbited by two smaller bodies. We distinguish between the target body with mass $m_t \ll M$, and the projectile body with mass $m_p \le m_t$. Note that in a co-rotating frame centered on the target, the target is at rest, and the projectile is seen to scatter off the target (see Fig.~\ref{fig:orbits}). 
The target's orbit around the central body is initially circular with semi-major axis $a_t$. The projectile body is also put on an initially circular orbit around the central body, in the same plane, but with semi-major axis $a_p = a_t + p$. 
Here, we define $p$ to be the impact parameter of our scattering experiment \citep[following][]{Petit86}, which can take on positive and negative values.  
A non-zero impact parameter causes the target and projectile bodies to have slightly different orbital periods. As a consequence, their separation gradually decreases in time, until it reaches a minimum. If this minimum is much larger than the Hill radius, then the result will be a single flyby. If the minimum is of order the Hill radius, then Jacobi captures start to manifest themselves. Since in the general case, the target and projectile bodies can have arbitrary mass ratios, we define the ``binary Hill radius'' as

\begin{equation}
\label{eq:binRH}
    R_{H} \equiv r \left( \frac{m_t + m_p}{3M} \right)^{\frac{1}{3}}, 
\end{equation}

\noindent with $r$ the orbital radius in the disc, and $m_t$ and $m_p$ the mass of the target and projectile body, respectively. In the limit of $m_p \rightarrow 0$, we retrieve Eq.~\eqref{eq:R_J}, which we will refer to as the target's Hill radius, or $R_{H,t}$ for short. 
To maximally resolve the gravitational focusing between the target and projectile, we separate them initially by 180 degrees in mean anomaly. In Fig.~\ref{fig:exp_setup}, we show two examples of projectile orbits, which cross the target's Hill radius; one for a positive impact parameter (in red), and one for a negative value (in orange). 
We distinguish between positive and negative impact parameters in order to determine potential asymmetries between approaching from the inside or outside, and consequences for prograde vs. retrograde capture. 

We choose two sets of masses based on two systems where Jacobi captures are thought to be effective: planetary discs and active galactic nuclei (AGN) discs. For the planetary case, we will adopt masses based on the Sun, Earth and Moon. The origin of the Moon has been under debate for decades \citep[see reviews by e.g.][]{Stevenson_1987,2016JGRE..121.1573B} and has been explained by models including i) fission from a fast spinning proto-Earth \citep[e.g.][]{1963JGR....68.1547W}, ii) co-accretion \citep[e.g.][]{2001EP&S...53..213M},  iii) tidal capture \citep[e.g.][]{1955ZA.....36..245G, 1955IrAJ....3..245O, 1967RSPSA.296..285L}, and iv) a Moon forming giant impact \citep[e.g.][]{Hartmann75}. The Moon’s low density and lack of Fe with respect to the solar/chondritic composition, opposes models describing the Moon’s co-formation near the proto-Earth, or any models assuming an intact capture of the Moon. Similarly to terrestrial rocks, lunar minerals have higher oxygen content than meteorites of the asteroid belt (or Mars), indicating that their silicate shells likely interacted in the past. These observations favour the giant impact theory, where a Moon-forming impactor called Theia followed a trajectory ending up in a collision with the proto-Earth, after which the Moon formed from the orbiting debris \citep[e.g.][]{2001Natur.412..708C, Canup_2008,  2012Sci...338.1052C, 2012Sci...338.1047C}. Moreover, a number of elements (e.g. Cr, Ti, Si, W; \citealt{1998GeCoA..62.2863L,2007Natur.447.1102G,2012NatGe...5..251Z}), show identical or close to identical \citep[e.g. O;][]{Cano_2020} stable isotopic composition to the values of the terrestrial mantle, indicating that the mantle of the Earth must have been largely, but not completely equilibrated with Theia \citep{Touboul_2007, Touboul_2015, Pahlevan_2014} and Theia formed at perhaps similar, but not completely identical heliocentric distance \citep{Cano_2020}. 
We will demonstrate that Jacobi captures can provide the dynamical background for both giant impacts and tidal captures. The latter is of interest as a proof of concept for other irregular moons in the Solar System. 

For the AGN case, we adopt masses based on Sagittarius A* \citep{2009ApJ...692.1075G, 2008ApJ...689.1044G} and GW190521 \citep{2020PhRvL.125j1102A}. This gravitational wave source was produced by rather massive black hole progenitors. Furthermore, the high eccentricity and spin parameters estimated for GW190521 are consistent with mergers in AGN discs \citep{2020PhRvL.124y1102G,2021ApJ...907L..20T,2021ApJ...908..194T}.
We define the mass ratio between the target and central body, $Q=m_t/M$, which takes on values of $3.0 \times 10^{-6}$ and $2.0 \times 10^{-5}$ for the planetary and AGN system respectively. We also define the mass ratio between the projectile and target bodies, $q=m_p / m_t\leq 1$, which takes on values of $1.2 \times 10^{-2}$ and $7.8 \times 10^{-1}$ for the planetary and AGN system respectively.

We adopt units in which the gravitational constant $G=1$, $M=1$ and $a_t=1$. The initial orbital period of the target body is thus $P_t \approx 2\pi$. 
The impact parameter is what we will vary systematically in units of the Hill radius.
We first vary $p$ from $1.0$ to $5.0\,R_{H,t}$ and from $-5.0$ to $-1.0\,R_{H,t}$ with a resolution of $1000$ evenly spaced sub-steps (i.e. $\Delta p = 0.004\,R_{\rm{H,t}}$, including the last value of $p$ resulting in 1001 simulations). This coarse sampling allows us to detect the interesting interval within which encounters closer than a Hill radius occur. We subsequently sample this smaller window again with $1000$ evenly spaced steps. This reveals multiple sub-windows within which more than 1 encounter occurs. In an iterative manner, we continue to detect the intervals of interesting sub-windows, which we then resolve with another 100 or 1000 sub-steps. We halted the iteration at the fourth level, at which point we performed a total of 187,460 simulations.

We integrate the three-body systems using the brute force N-body code \texttt{Brutus} \citep{2014ApJ...785L...3P, 2015ComAC...2....2B,2021PhRvD.104h3020B}. This code implements Bulirsch-Stoer iteration and arbitrary-precision arithmetic. We fix the Bulirsch-Stoer tolerance to $\epsilon = 10^{-6}$ and the word-length to $L_w = 128$ bits. These values are sufficient to smoothly integrate through all close encounters, even the near-collisional trajectories. An in-depth analysis of the Lyapunov time scale \citep[e.g.][]{ph421} of Jacobi captures and the effect of numerical divergence \citep[e.g.][]{1964ApJ...140..250M, 2014ApJ...785L...3P, 2018CNSNS..61..160P,2020MNRAS.493.3932B} is left for a future study.
Relativistic corrections and tidal effects are neglected in the simulations, but they are accounted for in the post-processing analysis in Sec.~\ref{sec:dissipative}.  

For each individual scattering experiment, we record the following quantities: 1) the duration, which we define as the total time that the projectile spends within a Hill radius from the target, 2) the number of close encounters between the projectile and target bodies, and 3) the separation and relative speed between the projectile and target bodies during each close encounter. 
We identify close encounters by locating the minima in the separation versus time curve, but only including those that fall within a Hill radius. The snapshot output interval occurs in tandem with the adaptive time steps of the integration, so that close encounters are sufficiently resolved.   
The stopping conditions are: 1) the true anomalies of the two bodies in their orbit around the central body, differ by more than 90 degrees and is increasing (e.g. large azimuthal angle in the orbital plane), or 2) a maximum of 100 close encounters is reached. In the first case, the two bodies have passed each other and are now far beyond the Hill radius. In the second case, the maximum of 100 encounters is chosen arbitrarily in order to limit computation time. This condition does not affect the main results of this paper.

\section{Jacobi captures in the conservative regime}\label{sec:result}

In this section we present our results for Jacobi captures without energy dissipation. In Sec.~\ref{sec:3.1} we present some illustrations of Jacobi captures with a varying number of close encounters. In Sec.~\ref{sec:3.2} we present the ``Jacobi spectra'', i.e. the number of encounters and closest separation as a function of impact parameter. In Sec.~\ref{sec:3.3}, we provide new empirical evidence suggesting that the phase space structure is self-similar, and we also present statistical results related to Jacobi captures. 

\subsection{Illustrations of Jacobi captures}\label{sec:3.1}

\begin{figure*}
\centering
\begin{tabular}{ccc}

\includegraphics[height=0.282\textwidth,width=0.33\textwidth]{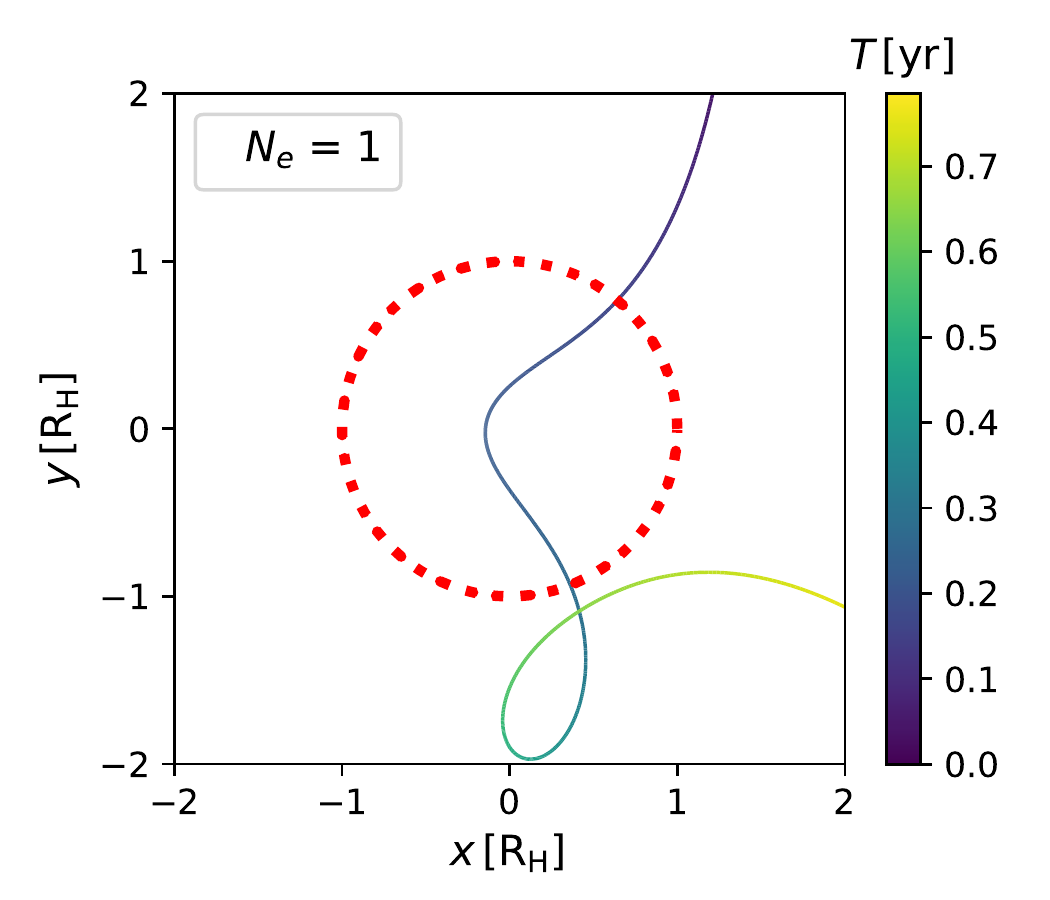} &
\includegraphics[height=0.282\textwidth,width=0.33\textwidth]{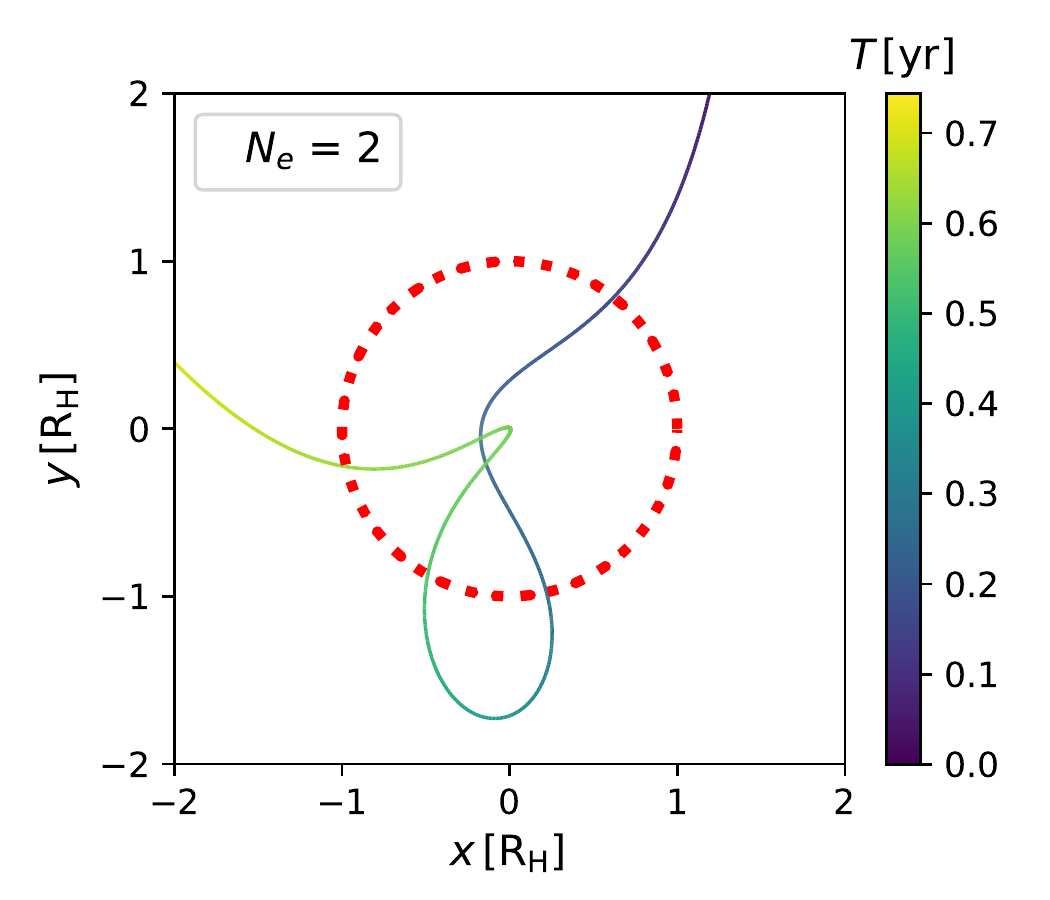} &
\includegraphics[height=0.282\textwidth,width=0.33\textwidth]{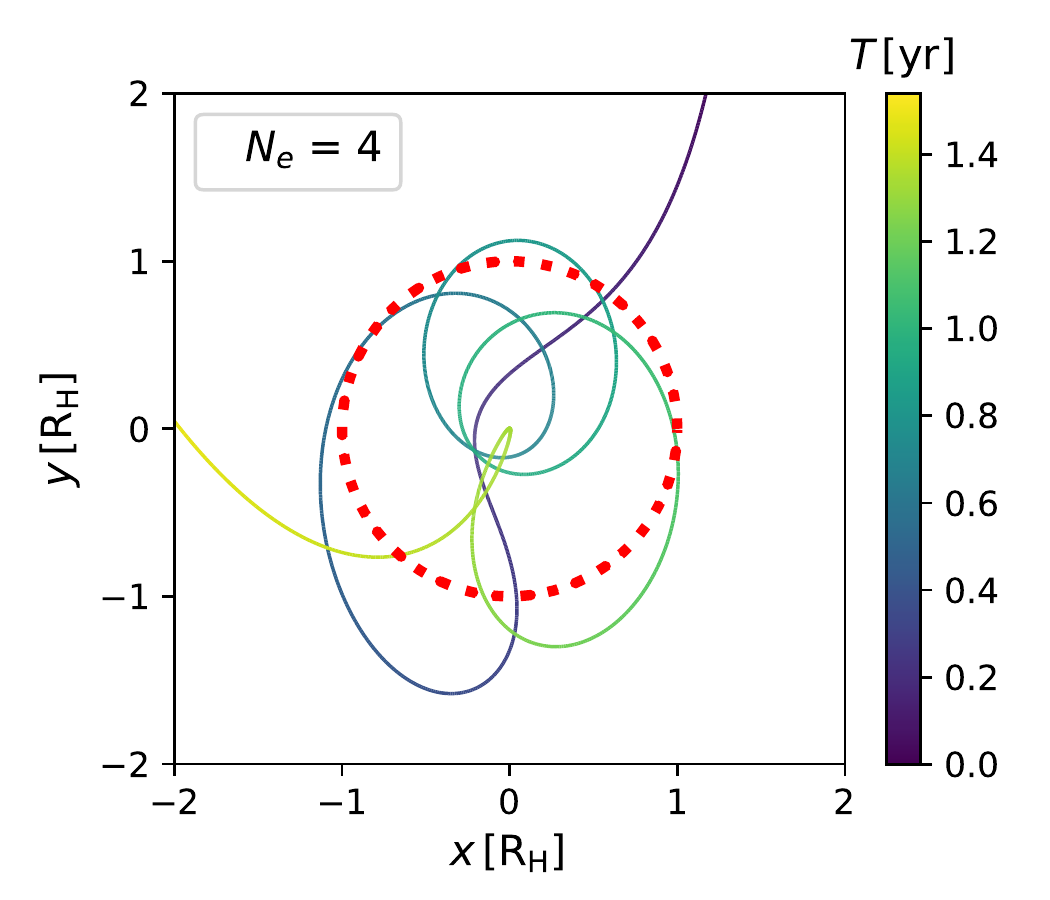} \\

\includegraphics[height=0.282\textwidth,width=0.33\textwidth]{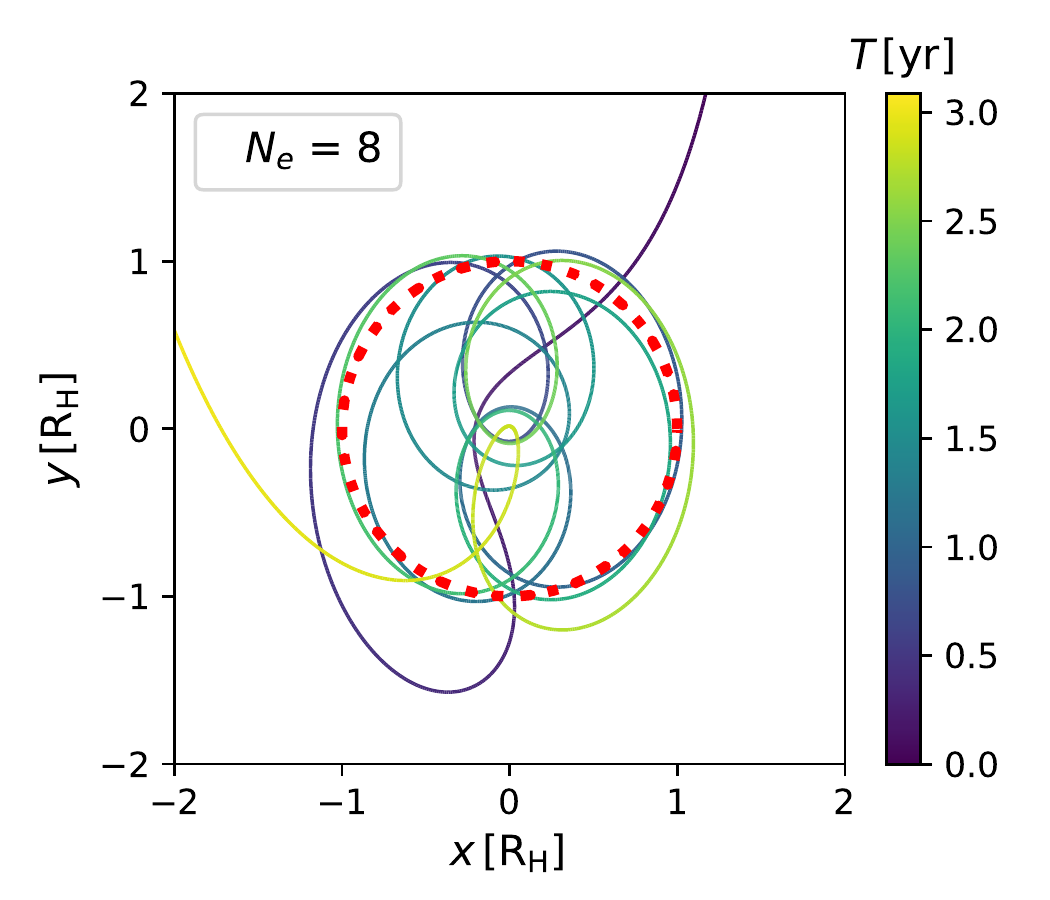} &
\includegraphics[height=0.282\textwidth,width=0.33\textwidth]{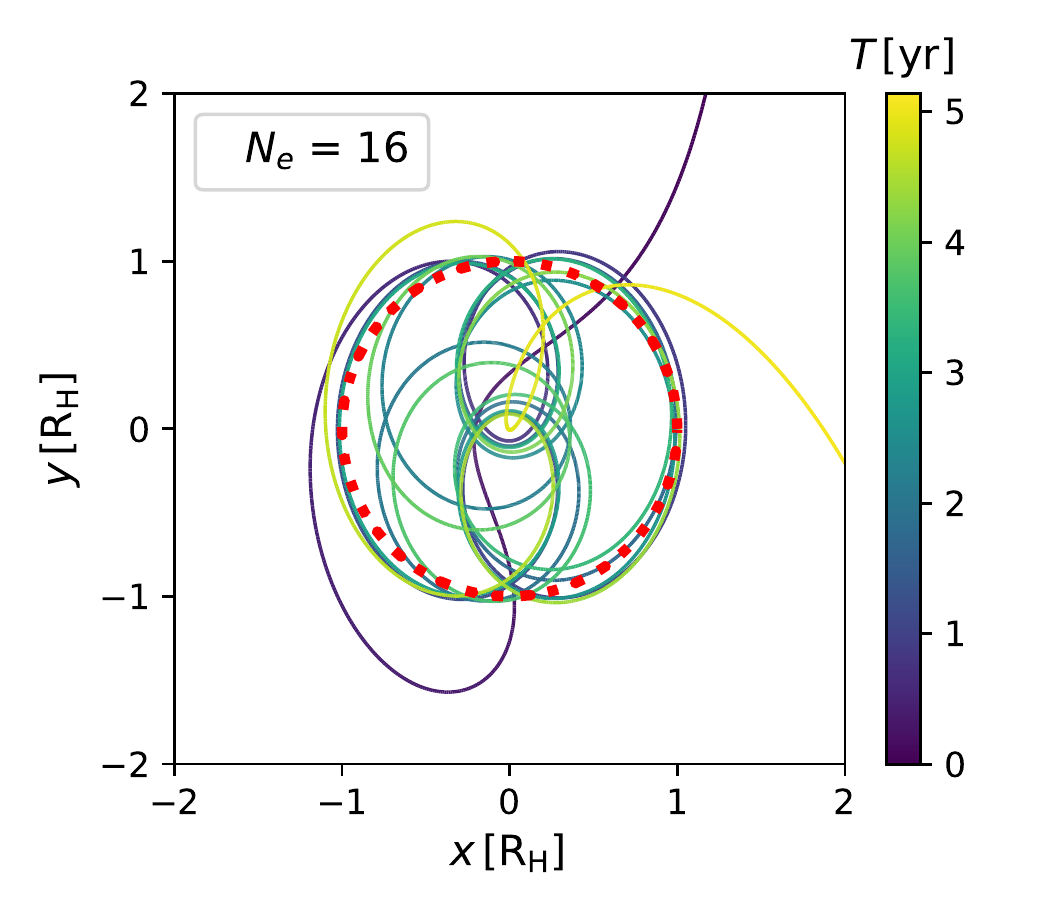} &
\includegraphics[height=0.282\textwidth,width=0.33\textwidth]{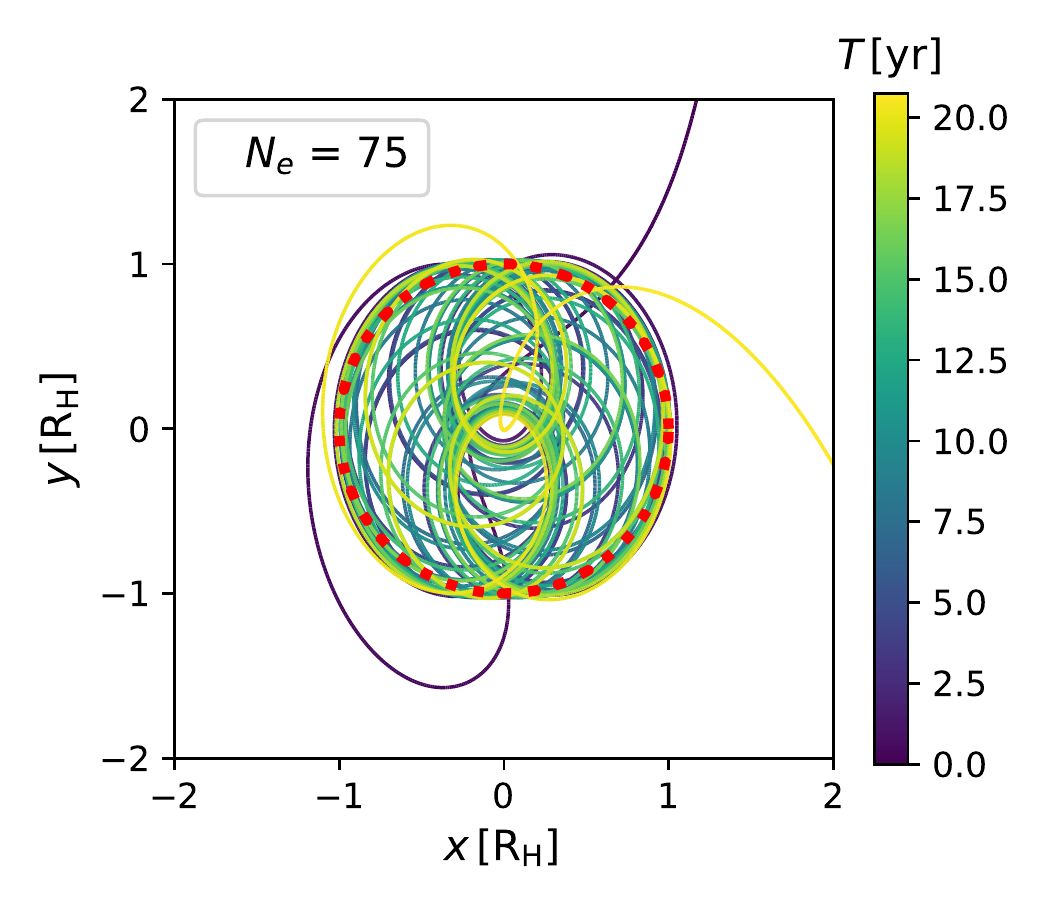} \\

\end{tabular}  
\caption{ Illustrations of Jacobi captures for the Sun-Earth-Moon system. The frame is at rest with respect to the target body (Earth) located at the origin, while the central body (Sun) is located towards the $-x$ direction. The projectile body (Moon) traces the orbit, which is colour coded according to time. The binary Hill radius is marked by the dotted circle. The number of close encounters, $N_e$, is given in the box.  }
\label{fig:orbits}
\end{figure*}

\begin{figure*}
\centering
\begin{tabular}{ccc}

\includegraphics[height=0.282\textwidth,width=0.33\textwidth]{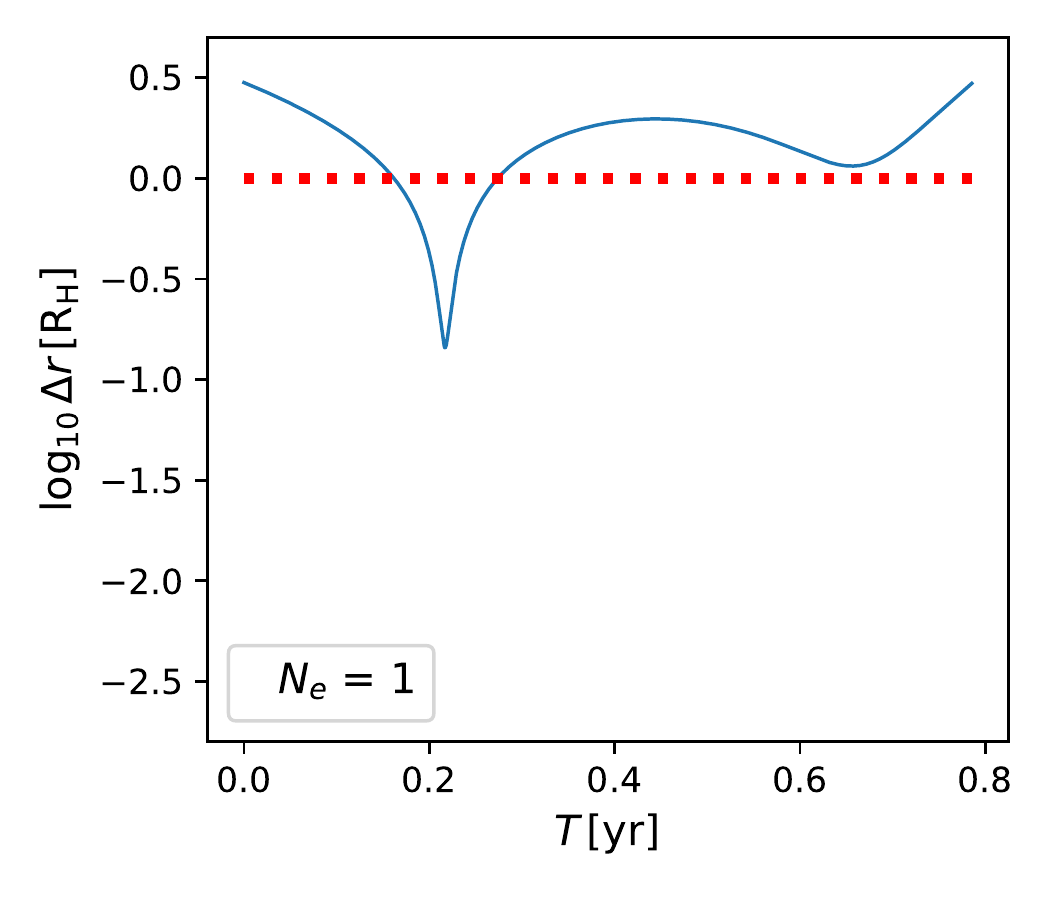} &
\includegraphics[height=0.282\textwidth,width=0.33\textwidth]{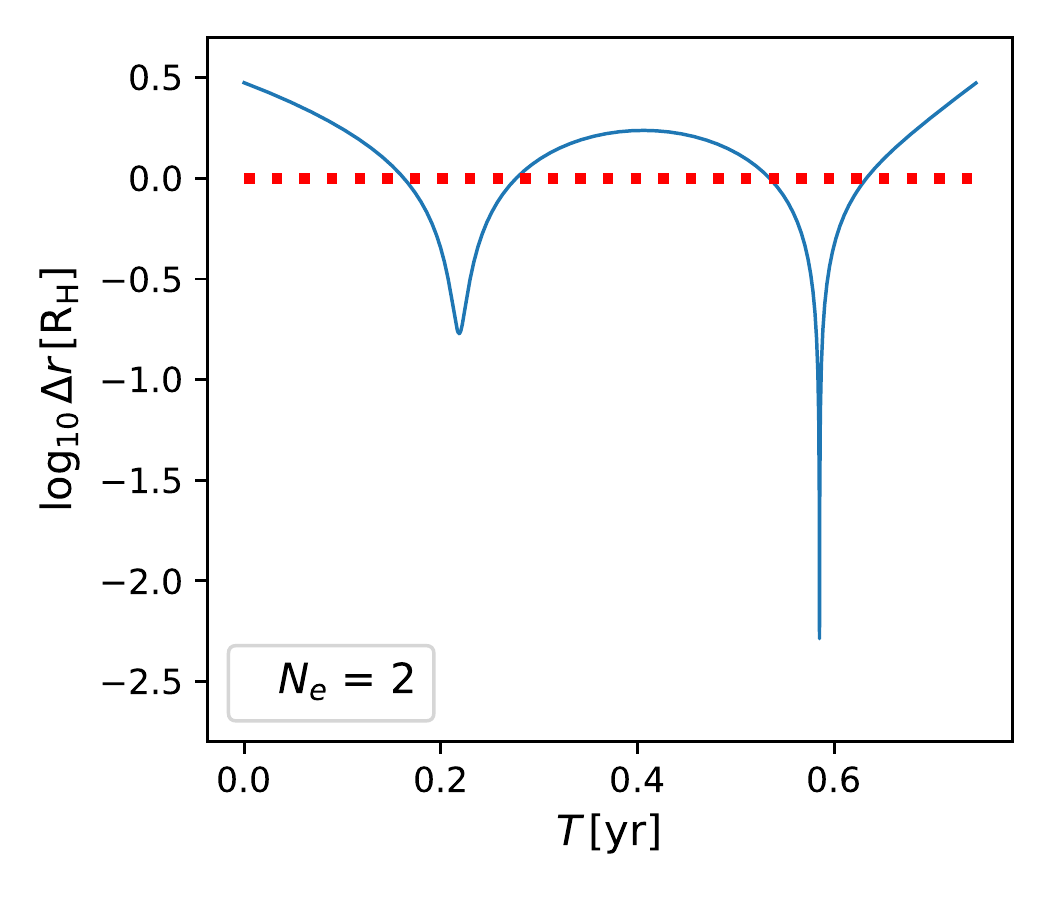} &
\includegraphics[height=0.282\textwidth,width=0.33\textwidth]{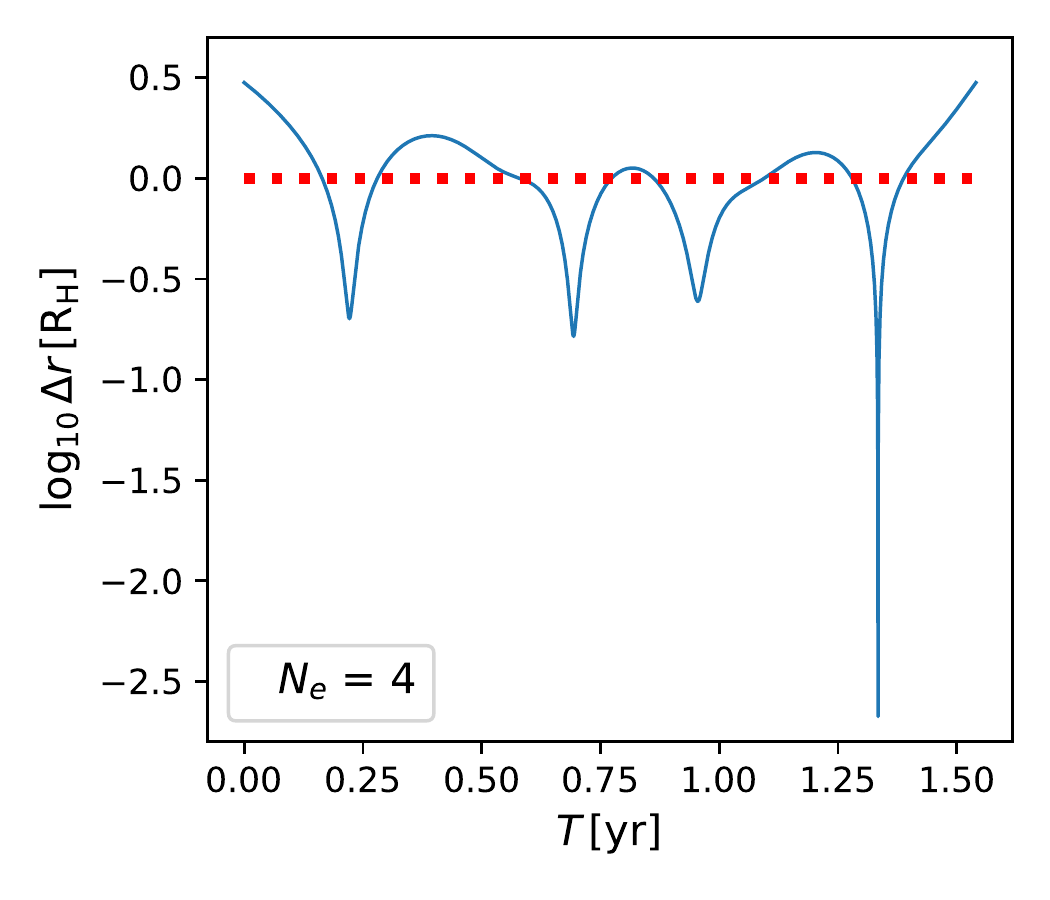} \\

\includegraphics[height=0.282\textwidth,width=0.33\textwidth]{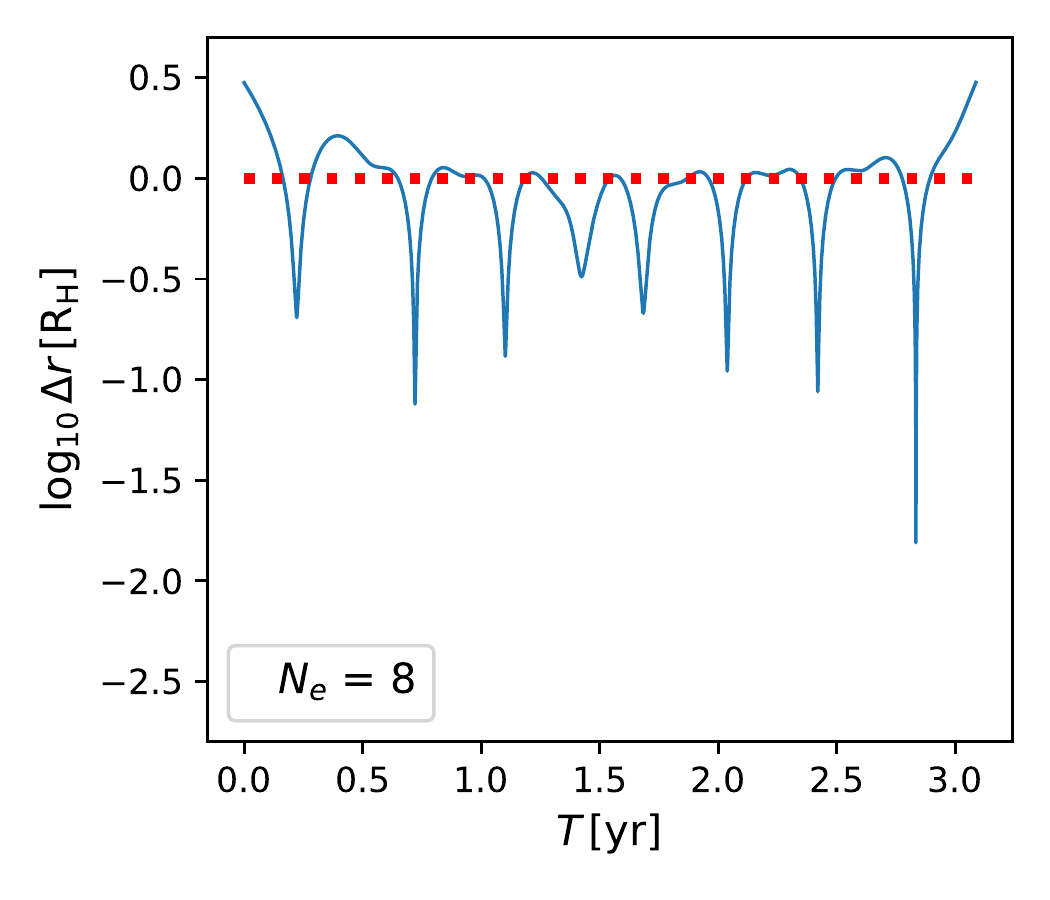} &
\includegraphics[height=0.282\textwidth,width=0.33\textwidth]{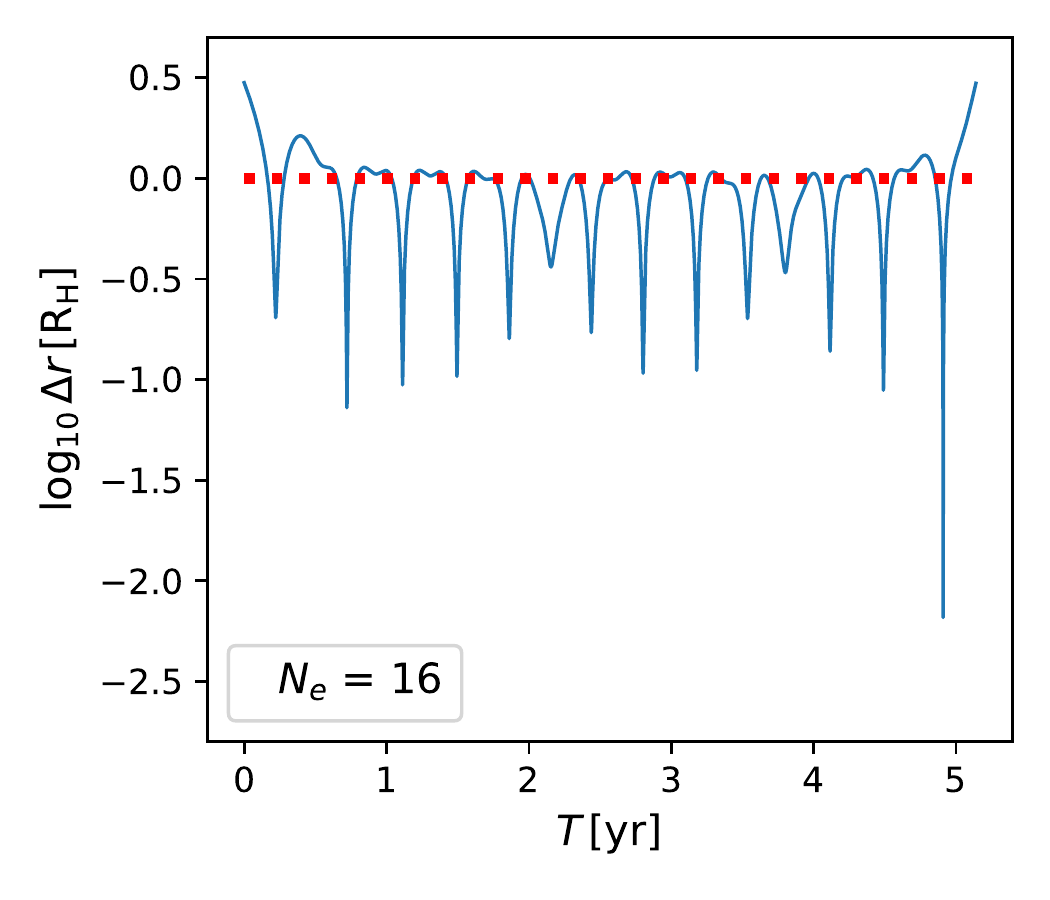} &
\includegraphics[height=0.282\textwidth,width=0.33\textwidth]{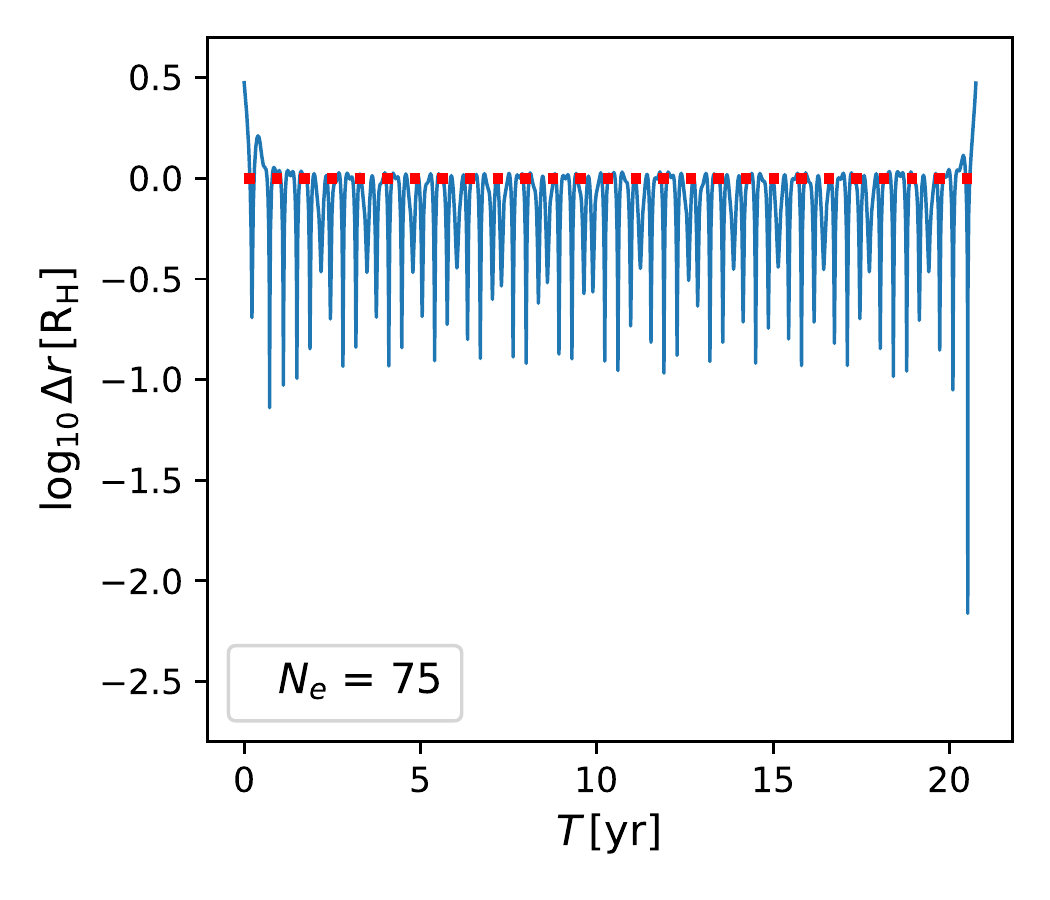} \\

\end{tabular}  
\caption{ Complementary to Fig.~\ref{fig:orbits}, we plot the time evolution of the separation between the projectile and target bodies.  }
\label{fig:dr_vs_t}
\end{figure*}

We consider the Sun-Earth-Moon initial condition described in Sec.~\ref{sec:methods}. We first consider an impact parameter of $p/R_{H,t}=1.92$. In the top left panel of Fig.~\ref{fig:orbits}, we visualise the orbit of the projectile body. The frame is centered 
on the target body, and rotated such that the central body (Sun) is on the x-axis in the negative direction. We observe the projectile body entering the Hill radius, after which a single close encounter occurs. 
In the corresponding top left panel of Fig.~\ref{fig:dr_vs_t}, we plot the separation between the projectile and target bodies for the same simulation. We confirm one minimum below the Hill radius threshold. 

Next, we slightly perturb the impact parameter to $p/R_{H,t}=1.91$. We find that this value results in a Jacobi capture with two close encounters (top, middle panels of Figs.~\ref{fig:orbits} and \ref{fig:dr_vs_t}).  
Repeating the simulations for slightly different initial conditions, we find that for $p/R_{H,t}=1.8988$ we obtain a solution with 4 close encounters. For $p/R_{H,t}=1.89799$ we get 8 close encounters, for $p/R_{H,t}=1.8979621$ we get 16, while for $p/R_{H,t}=1.8979619$ we obtain a relatively long lived Jacobi capture with 75 close encounters. It is such prolonged Jacobi captures which provide the dynamical background for close encounters, which may lead to binary formation if a source of energy dissipation is taken into account.
Furthermore, this empirical sequence demonstrates that there is a sensitive dependence on initial conditions, which we characterise further in the next two sections. For our second system based on black holes in AGN discs, we also obtain Jacobi captures of various duration, which we will show in the next section.

\subsection{Jacobi capture spectra}\label{sec:3.2}

\begin{figure*}
\centering
\begin{tabular}{c}
\includegraphics[height=0.475\textwidth,width=0.95\textwidth]{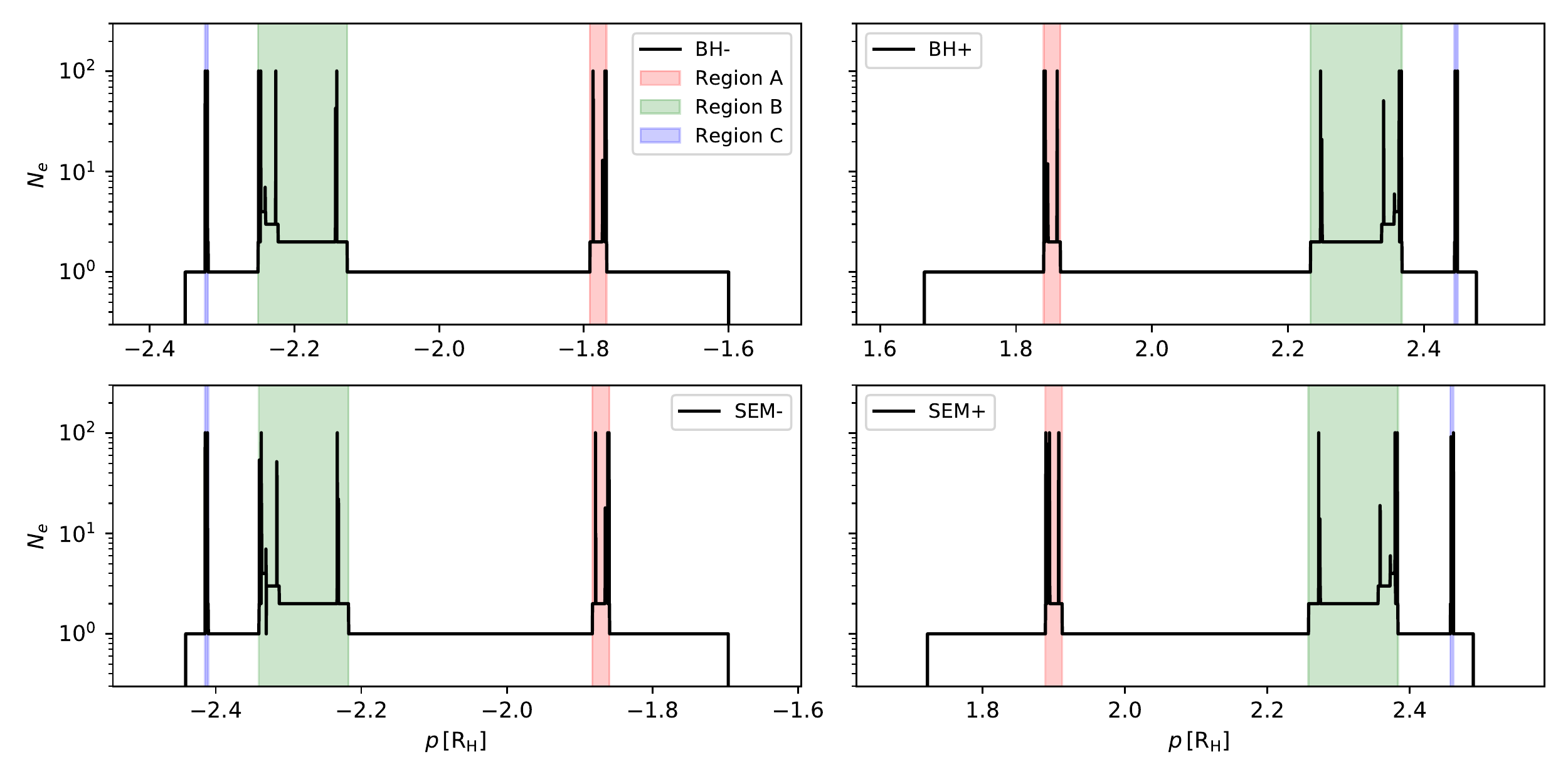} \\
\end{tabular}  
\caption{ Number of close encounters, $N_e$, as a function of impact parameter, $p$, which is in units of the binary Hill radius, $R_H$. The top row panels give the ``spectra'' for the black hole (BH) system, and the lower row for the Sun-Earth-Moon (SEM) system. We also differentiate between positive (right column) and negative (left column) impact parameters. We re-scaled the horizontal axis of each panel in order to align the spectra. The colour bars mark three characteristic regions in which Jacobi captures occur.   }
\label{fig:spectra_ne}
\end{figure*}

\begin{figure*}
\centering
\begin{tabular}{c}
\includegraphics[height=0.475\textwidth,width=0.95\textwidth]{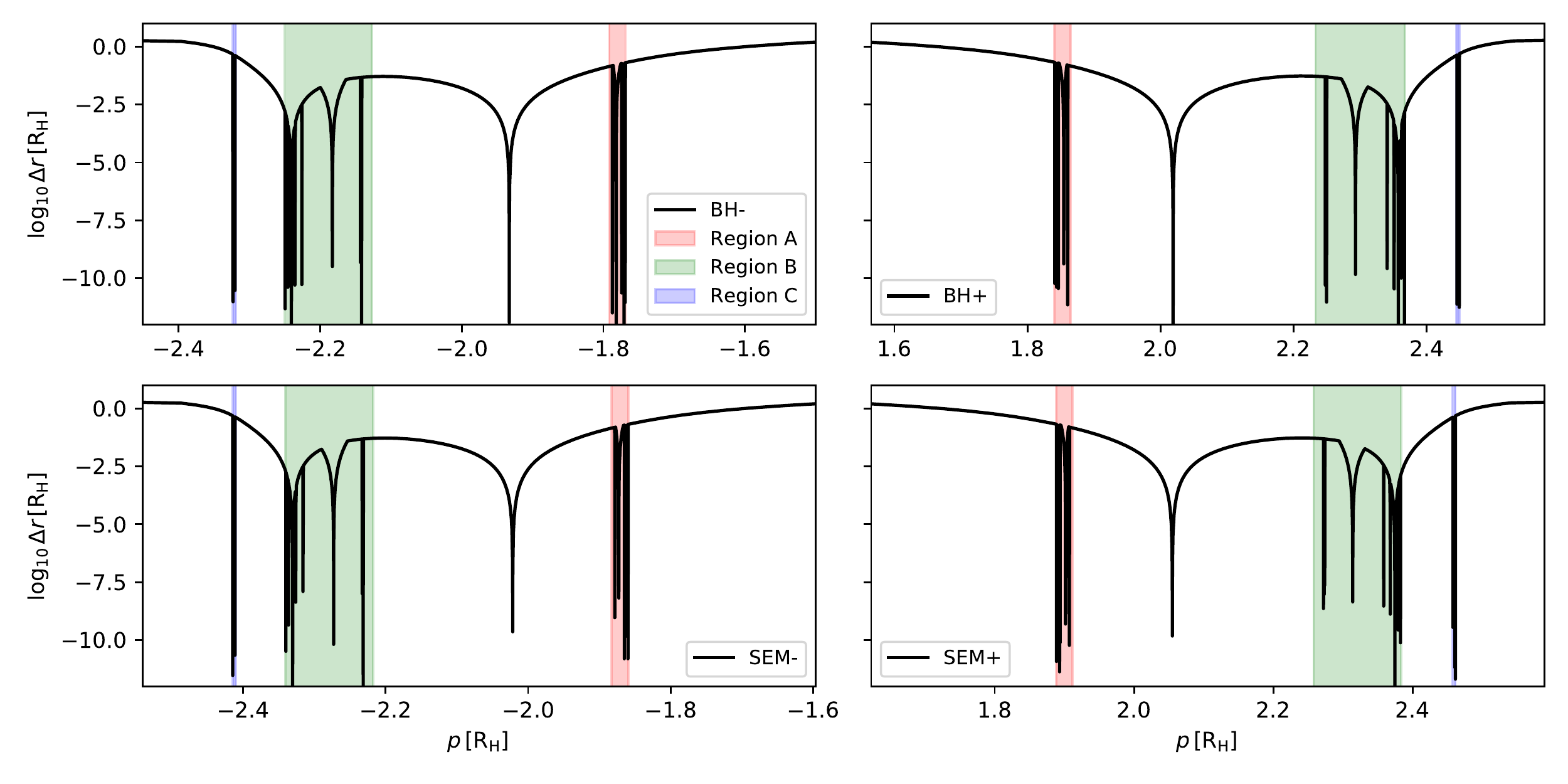} \\
\end{tabular}  
\caption{ Similar to Fig.~\ref{fig:spectra_ne}, but for the closest approach between the target and projectile bodies, $\Delta r$, in units of the binary Hill radius, $R_H$. The colour bars are the same as in Fig.~\ref{fig:spectra_ne}. }
\label{fig:spectra_dr}
\end{figure*}

\begin{figure}
\centering
\begin{tabular}{c}
\includegraphics[height=0.384\textwidth,width=0.48\textwidth]{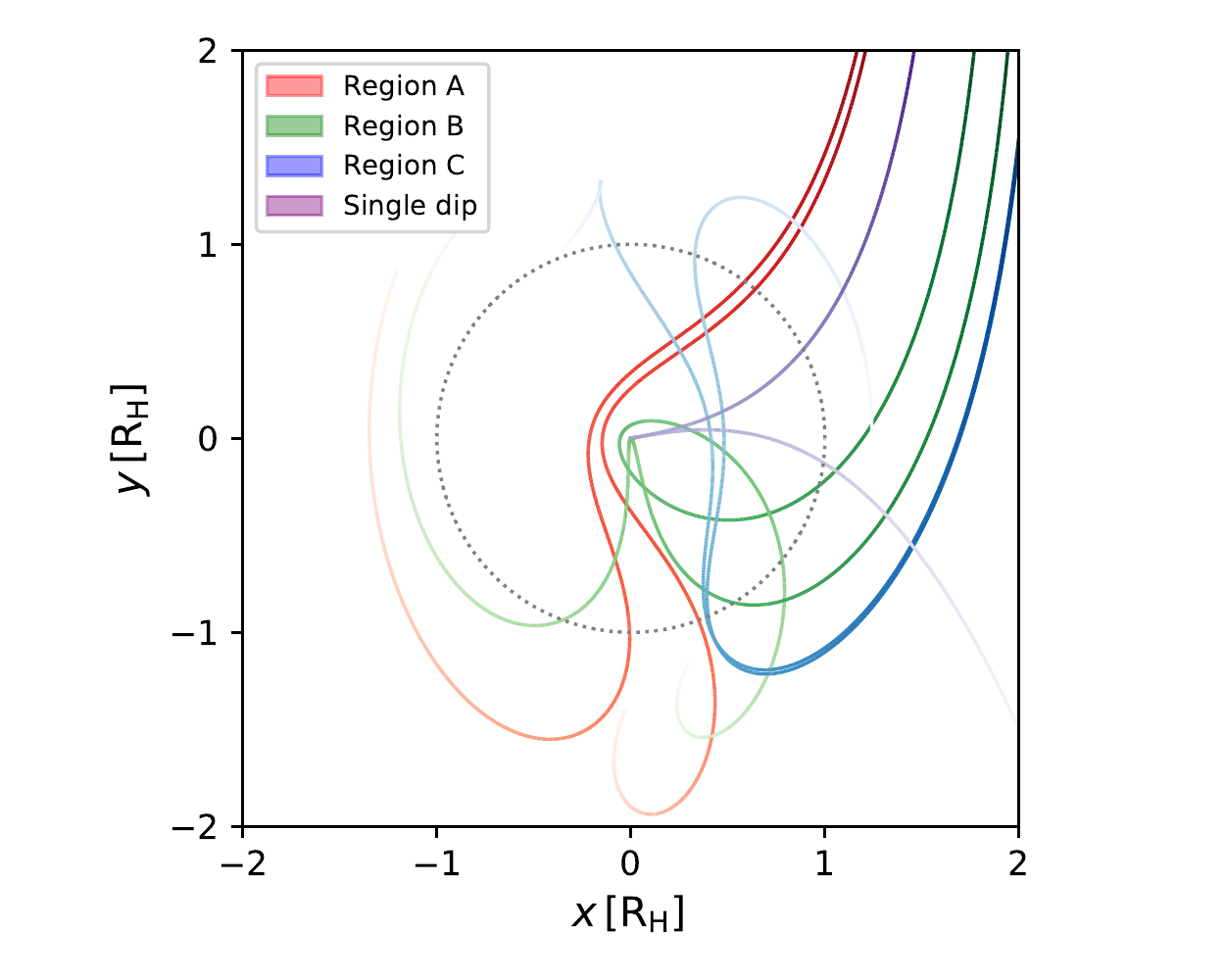} \\
\end{tabular}  
\caption{ Illustration of the orbits belonging to the edges of the three sub-regions as well as the single dip feature in the spectra of  Figs.~\ref{fig:spectra_ne}-\ref{fig:spectra_dr}. The same colour code is adopted here. The data is for the Sun-Earth-Moon system with positive impact parameters.  
The projectile body enters the plot from the upper right and moves downwards initially. 
}
\label{fig:orbit_types}
\end{figure}

\begin{figure*}
\centering
\begin{tabular}{c}
\includegraphics[height=0.665\textwidth,width=0.95\textwidth]{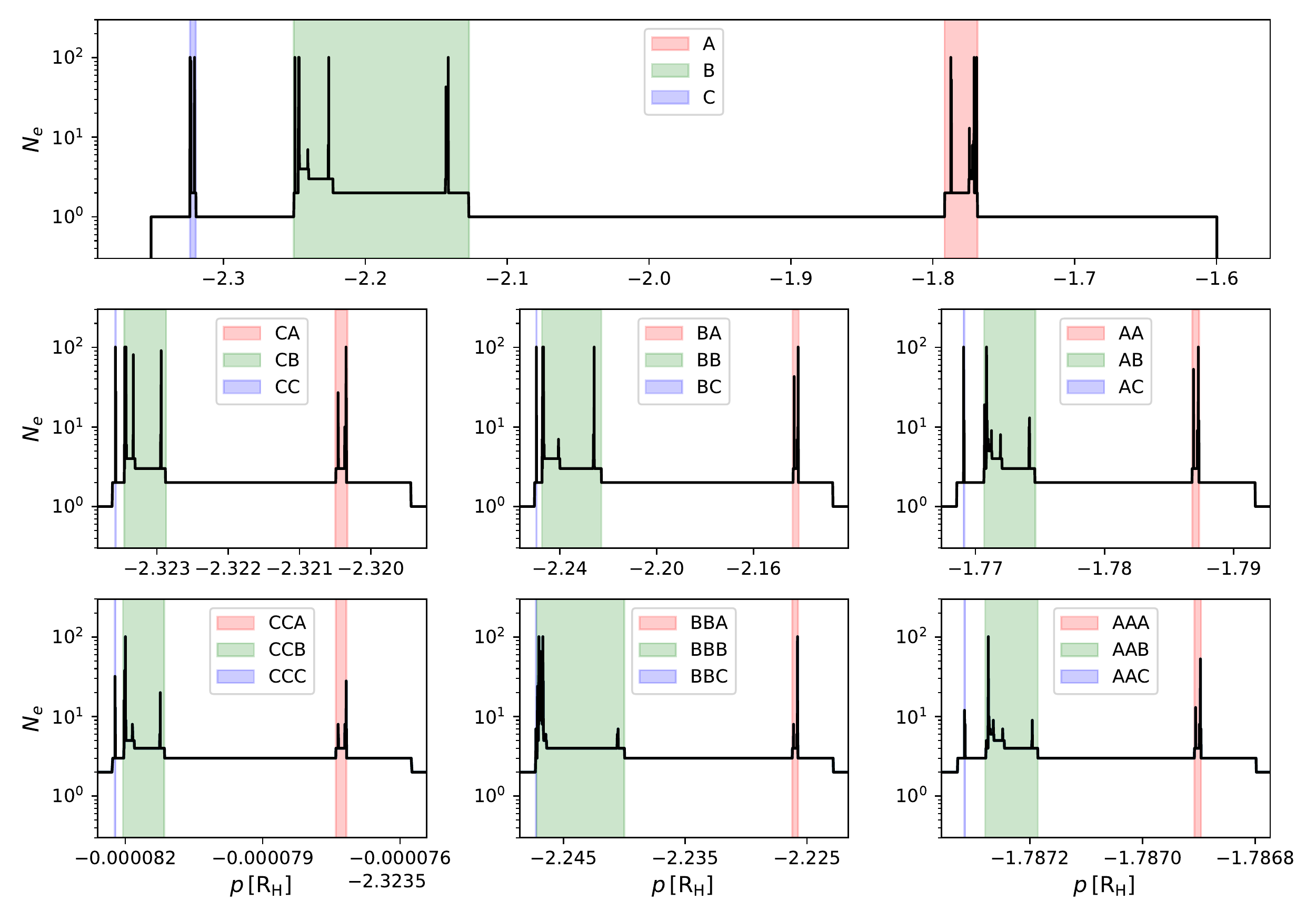} \\
\end{tabular}  
\caption{ Number of close encounters, $N_e$, as a function of impact parameter, $p$, which is in units of the binary Hill radius, $R_H$. The data is for the black hole (BH) system. In the top row, we show the complete spectrum, with the three different islands marked by the shaded regions. In the subsequent rows, we zoom in on a shaded region in the panel above it. The first letters denote the zoomed-in regions of the spectra above, while the last letter denotes the islands in the current zoom-in. Note that in some panels, the x-axis is reversed in order to align the spectra. }
\label{fig:self_similar_ne}
\end{figure*}

Here we consider two statistical properties of a Jacobi capture: the total number of close encounters, $N_e$, and the closest separation between the target and projectile bodies during the entire interaction, $\Delta r$. We plot these quantities as a function of impact parameter in Figs.~\ref{fig:spectra_ne} and \ref{fig:spectra_dr}.  

In Fig.~\ref{fig:spectra_ne}, we observe the broad base line for which only a single close encounter occurs within the Hill radius, ``a flyby''. As expected, the largest portion of this impact parameter space results in a flyby. However, superposed are three distinct ``islands'' (marked by the colour bars) in which the number of encounters is higher. Sharp spikes are observed reaching up to 100 encounters, which is the allowed maximum in our experiment. 
Without such a constraint and with a sufficient resolution, these spikes are expected to grow to infinity, representing increasingly long lived Jacobi captures.
It is clear from these ``Jacobi spectra'' that there is an underlying structure in the parameter space of initial conditions which leads to multiple encounters. We will discuss this further in the next section.

In Fig.~\ref{fig:spectra_dr}, we plot the closest separation, which is a continuous variable. The baseline is therefore not as straight as in Fig.~\ref{fig:spectra_ne}, but the overall structure is the same, i.e. the dips belonging to close encounter trajectories 
are confined to the same three characteristic islands. There is one exception, which is the deep dip
outside any colour bar (towards the middle of each spectrum), which corresponds to a single close flyby. Most close encounter trajectories however, are found within the three islands which are associated with Jacobi captures, i.e. interactions with more than one encounter. 

In order to gain an intuitive understanding on the origin of these three islands, we take the impact parameters at the edges of each island (for the Sun-Earth-Moon system with positive impact parameters), as well as for the single dip feature in the spectrum. We plot their associated orbital evolution in Fig.~\ref{fig:orbit_types}. The coordinate frame is the same as in Fig.~\ref{fig:orbits}. We observe that projectile bodies belonging to region $A$ experience strong gravitational focusing prior to the close passage with the target. As a consequence, they are deflected inwards towards the central body, and encounter the target from the inside. This results in prograde encounters, i.e. same direction as their orbits around the central body (counter-clockwise in this plot). Projectile bodies in region $C$
are initially further away. Consequently, gravitational focusing is delayed until the projectile body is trailing the target body.  
Due to the gravitational pull of the target body however, the projectile body accelerates to overtake the target again from the outside. These encounters are also prograde.  
Projectile bodies in region $B$ approach the target more closely. The orbit at the inner edge encounters the target in a retrograde manner, while the orbit at the outer edge has a prograde encounter. 
Region $B$ thus exhibits a transition between prograde and retrograde encounters. We return to this observation in our discussion of Fig.~\ref{fig:pro_vs_ret}. 
Hence, the three orbit families are differentiated in their type of encounter with the target, i.e. from the inside (prograde), from the outside (prograde) or approximately head-on (prograde and retrograde). Trajectories corresponding to the single dip feature in the spectrum also lead to very close encounters. The center of the dip corresponds to an exact head-on collision, while slightly offset trajectories result in either prograde or retrograde encounters. 

For bodies with large sizes in units of their Hill radius, all four orbit families can result in giant impacts. However, if the sizes of the bodies are reduced, there will come a point where close approaches in regions $A$ and $C$ are too wide for impacts. In that case, giant impacts are most likely to originate from trajectories in region $B$ or the range of impact parameters within the single dip feature. 

\begin{figure*}
\centering
\begin{tabular}{c}
\includegraphics[height=0.54\textwidth,width=0.82\textwidth]{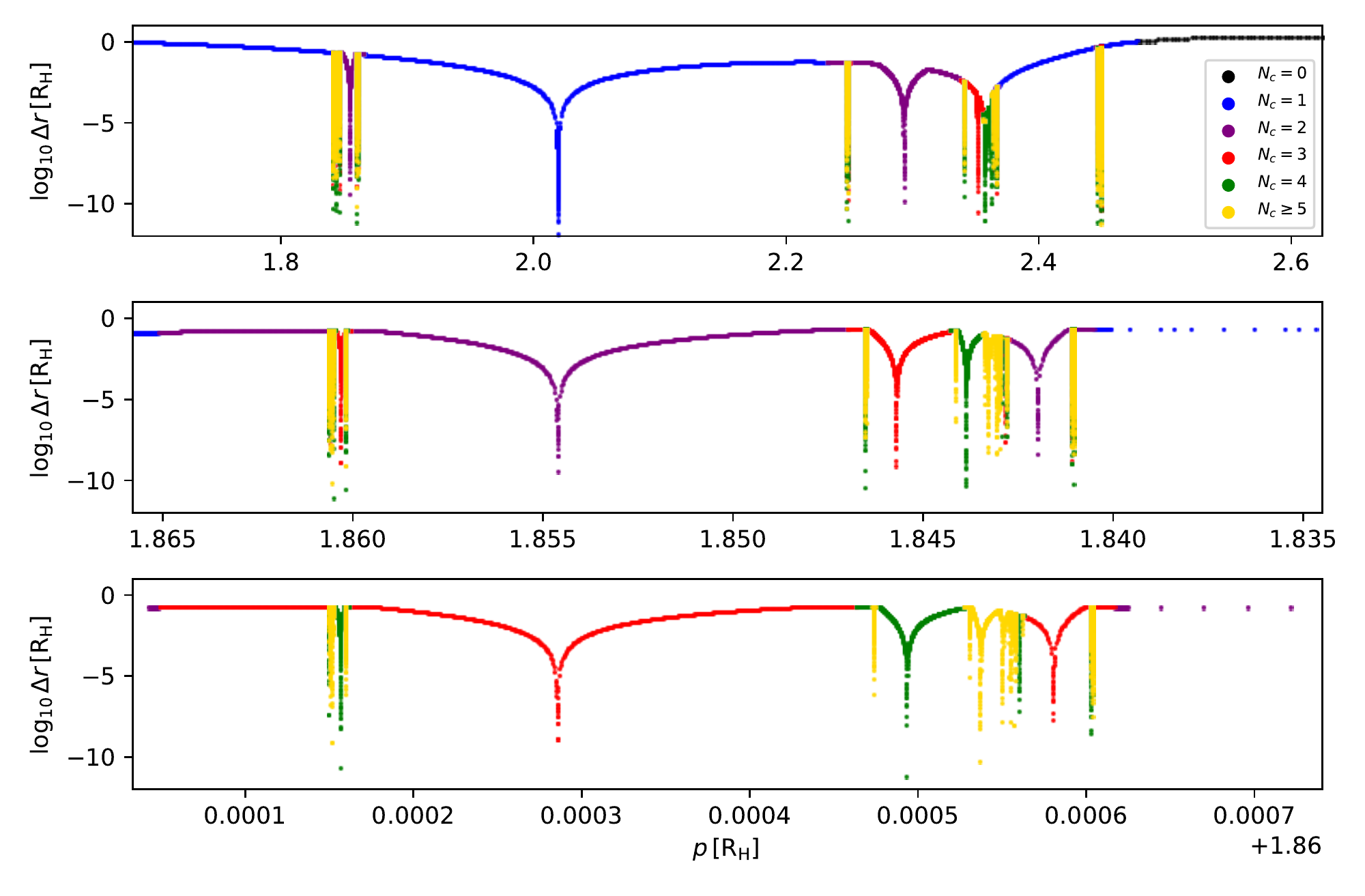} \\
\end{tabular}  
\caption{ Self-similarity in the spectrum of closest encounters for the BH system with positive impact parameters. The top panel is the same as the top right panel of Fig.~\ref{fig:spectra_dr}. Here, we colour code the data by the index, $N_c$, of the closest encounter. 
Note that we only consider close encounters within the Hill radius, e.g. $N_c=0$ corresponds to no encounters within the Hill radius.
The panels below are a zoom in of region $A$ in the panel above it. }
\label{fig:self_similar_dr}
\end{figure*}

\begin{figure*}
\centering
\begin{tabular}{c}
\includegraphics[height=0.54\textwidth,width=0.82\textwidth]{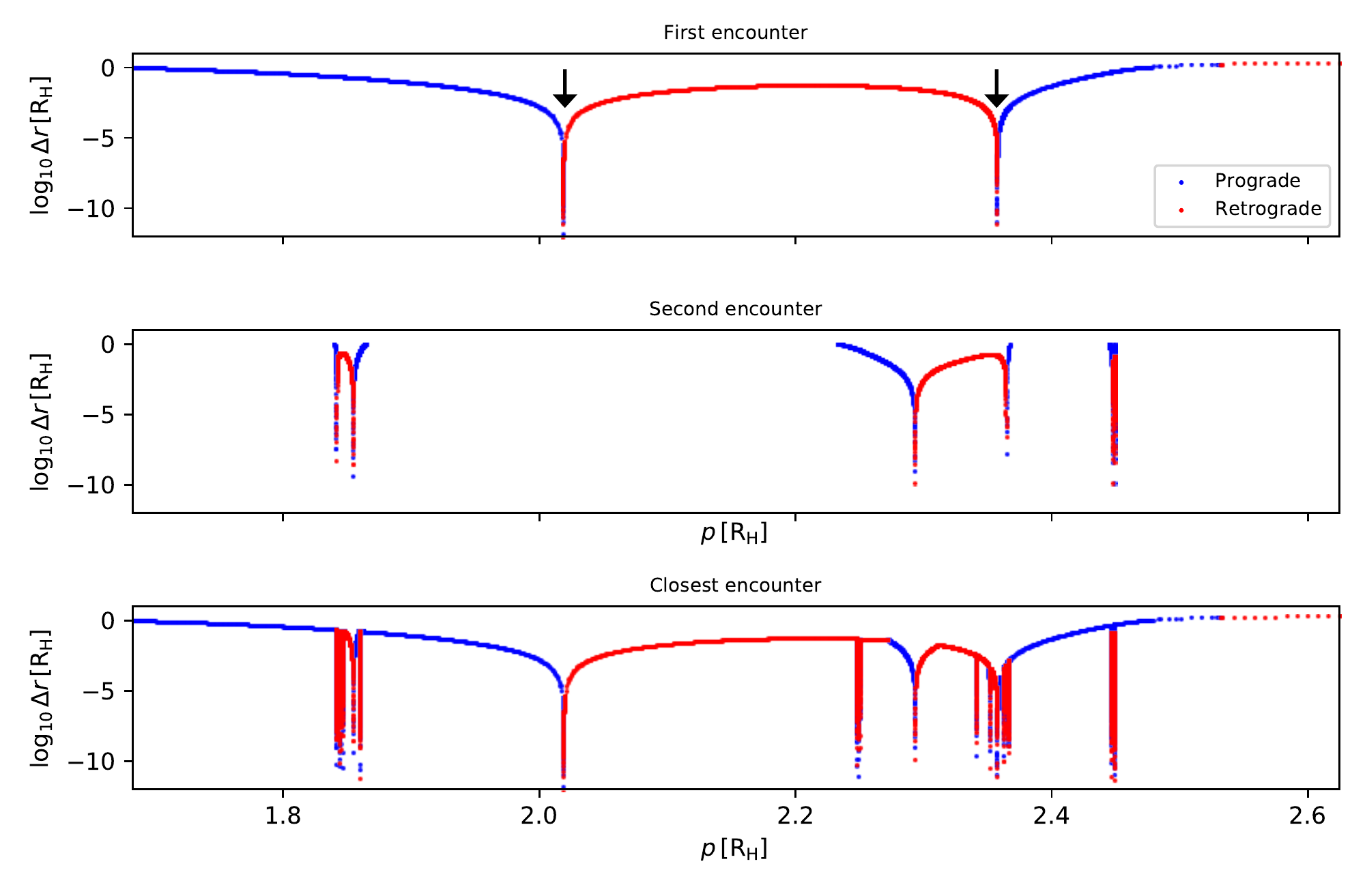} \\
\end{tabular}  
\caption{For the same data set as the top panel of Fig.~\ref{fig:self_similar_dr}, we plot the pericenter distance between the target and projectile body during their first encounter (top panel). We colour code the data by the orientation of the encounter: prograde vs. retrograde with respect to their orbits around the central body. The two arrows mark the head-on collisional trajectories of the first encounter, which correspond to transitions between prograde and retrograde encounters. The middle panel shows the impact parameters on the same scale leading to a second close encounter.
We observe three distinct regions in impact parameter space, where a second close encounter occurs. Each of them exhibits a self-similar structure, including two head-on collisional trajectories. The bottom panel shows the spectrum for the closest encounter (regardless of which close encounter it is in succession) during the entire Jacobi capture (same as top panel of Fig.~\ref{fig:self_similar_dr}).   }
\label{fig:pro_vs_ret}
\end{figure*}

\subsection{Self-similarity, fractal dimension and statistical properties of Jacobi captures}\label{sec:3.3}

\begin{figure}
\centering
\begin{tabular}{c}
\includegraphics[height=0.384\textwidth,width=0.48\textwidth]{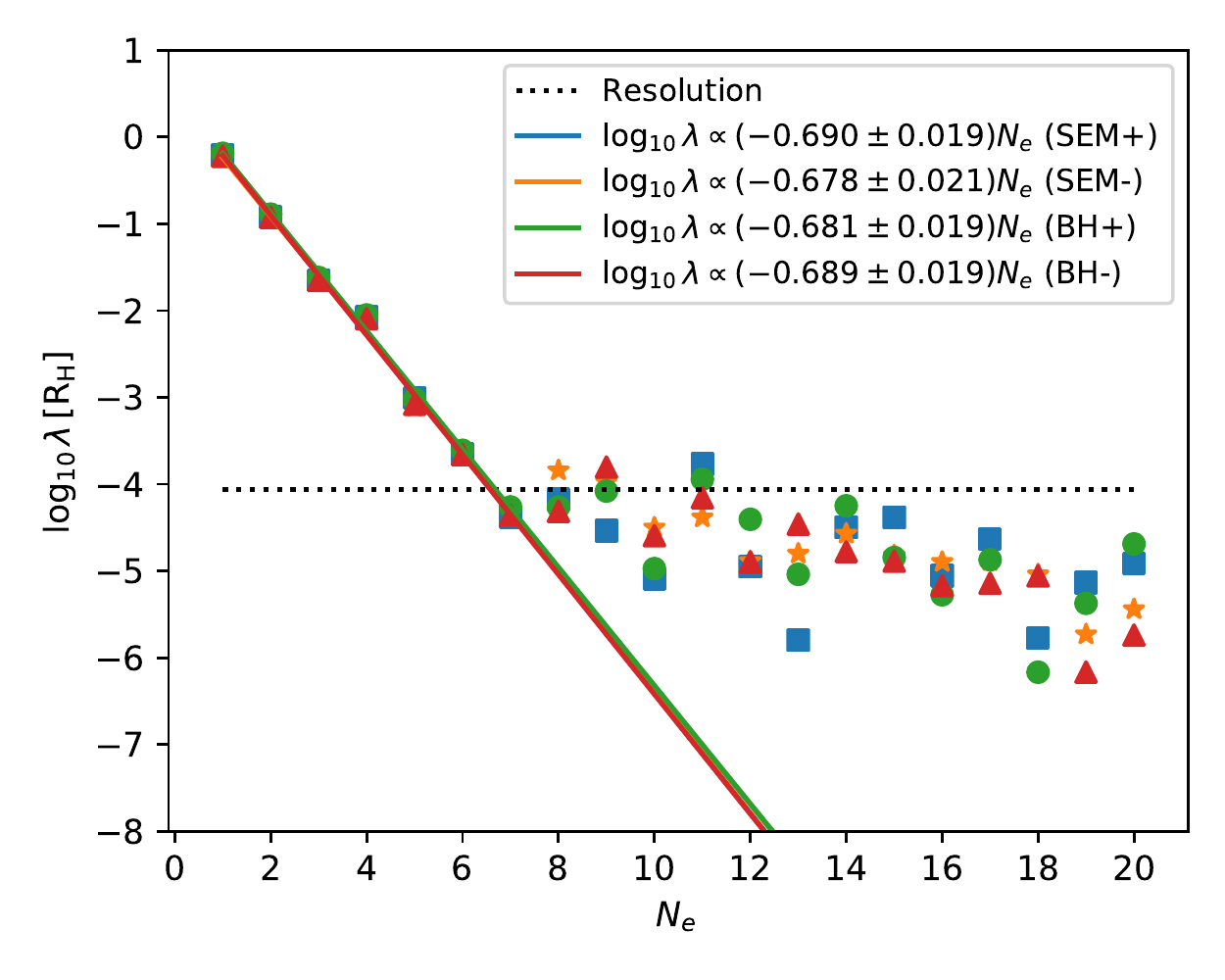} \\
\end{tabular}  
\caption{ Line section (i.e. 1D cross section) profile for the number of close encounters, $N_e$. Up until the average resolution of the spectrum (dotted line), we fit an exponential decay. }
\label{fig:lambda_vs_ne}
\end{figure}

As described in Sec.~\ref{sec:methods}, we iteratively sample the phase space of impact parameters in order to resolve the substructures in the Jacobi spectra. This allows us to zoom-in onto the three islands and study their local spectral properties. In Fig.~\ref{fig:self_similar_ne}, we replot the spectrum for the number of close encounters for the black hole system with negative impact parameters (top panel). The three panels below are a zoom-in of regions $A$, $B$ and $C$ in the panel above (represented by the first letter in the legend). We observe that the spectra are all self-similar, each reproducing their own three islands (see second letter in the legend). The baseline however is raised from a single encounter to two encounters. In the bottom row, we present the spectra for yet another zoom-in of an island in the panel above (again labelled by the letters in the legend). Although small differences start to appear due to numerical artefacts, we can still distinguish the three islands, and the raised baseline to three encounters. This sequence of zoom-ins demonstrates that the Jacobi spectrum for the number of close encounters is expected to be self-similar. 

We perform a similar exercise for the closest encounter spectrum belonging to the black hole system with positive impact parameters. In Fig.~\ref{fig:self_similar_dr}, we replot the closest encounter spectrum from Fig.~\ref{fig:spectra_dr}, but this time we colour code the data by the index of the encounter that was the closest one. For example, in the top panel, we confirm that the single dip feature outside of region $A$, $B$ and $C$, corresponds to the first (and only) close encounter (marked by the colour blue). In each panel below, we zoom in on region $A$ of the panel above it. We observe that each zoom-in reproduces the spectrum above it to a large degree. Not only the three separate islands are reproduced, but also the single dip feature, which is delayed to one later encounter in each zoom-in. Hence, the self-similarity is also evident in the Jacobi spectra for the closest encounters. 

In Fig.~\ref{fig:pro_vs_ret}, we further decompose the Jacobi spectrum of the closest encounters. Rather than focusing on the closest encounter of the entire Jacobi capture, we start by considering the pericenter distance of the first encounter (top panel of Fig.~\ref{fig:pro_vs_ret}). We observe two dips corresponding to two head-on, collisional trajectories (marked by the arrows). The existence of these two orbits was already touched upon in the discussion of Fig.~\ref{fig:orbit_types}. In particular, the collisional trajectory with the smaller impact parameter corresponds to the single dip feature of the Jacobi spectrum in Fig.~\ref{fig:spectra_dr}, while the other one falls within region $B$. We colour code the data according to whether the first encounter is prograde or retrograde. We confirm that both head-on collisional trajectories serve as transition points between prograde and retrograde encounters. Next, we focus on the pericenter distance of the second encounter (middle panel of Fig.~\ref{fig:pro_vs_ret}). We only observe data points inside the three sub-regions defined previously in Figs.~\ref{fig:spectra_ne} and \ref{fig:spectra_dr}. Each of them exhibits their own set of two head-on collisional trajectories (we verified this for region $C$ after zooming in). The impact parameters in between these two transition points correspond to retrograde encounters, while the outer parts are prograde. This structure appears to be self-similar among the three sub-regions, as well as among the different orders of the encounter. Furthermore, focusing on the impact parameter near $p=2.25\,\rm{R_H}$, we observe that the first encounter is retrograde, while the second encounter is prograde. Hence, this demonstrates that during a Jacobi capture, a projectile body can change its orbital orientation from one encounter to the next. 
The full spectrum for the closest encounter during the entire Jacobi capture (bottom panel of Fig~\ref{fig:pro_vs_ret}) can be thought of as being constructed by stacking up the head-on collisional trajectories of increasingly higher-order encounters. 

The self-similarity of the Jacobi spectra hints towards a fractal phase space structure. This was suggested before by e.g. \citet{Petit86}, who explain its origin by a hierarchical build up of higher-order transitions during which a system crosses multiple unstable, periodic orbits. Our detailed numerical results of the first few zoom-in levels provide new empirical evidence for such a fractal structure. By assuming that these results can be extrapolated to even higher levels, we are able to estimate the associated fractal dimension from the constant zoom-in factor derived from our spectra. We start by considering the discrete spectra related to the number of encounters (Figs.~\ref{fig:spectra_ne} and \ref{fig:self_similar_ne}).
The base level, i.e. the flat plateau belonging to a single encounter, has three superposed islands. These islands are directly related to the three orbit families in Fig.~\ref{fig:orbit_types}. Each individual island in the spectrum is elevated to the next level with a baseline of two encounters. Next, each of the  three sub-islands has their own set of three ``subsub-islands'', which are elevated to the next level with a baseline of three encounters. 
The observation that each zoom-in level reveals three smaller islands, is a demonstration of the fact that a system can switch between the three different orbital families. For example, in Fig.~\ref{fig:orbit_types}, the first close encounter might be within region A. If the outgoing orbit manages to turn around for the next close encounter, it might do that within each of the three regions A, B or C. Hence, prolonged Jacobi captures can be viewed as random permutations between regions A, B and C (see Sec.~\ref{sec:physical_picture}). We also observed in Fig.~\ref{fig:pro_vs_ret}, that subsequent close encounters can switch between prograde and retrograde. These are both manifestations of the transitions discussed by \citet{Petit86}.
If we were to extrapolate to even higher zoom-in levels, then the number of islands would increase exponentially, i.e. 
\begin{equation}
    N_{\rm{islands}} = 3^n,
\end{equation}

\noindent with $n$ the zoom-in level, and the base number 3 represents the number of islands per level, i.e. the three orbit families visualised in Fig.~\ref{fig:orbit_types}.
Furthermore, the baseline spans a certain range in impact parameters, which we define as the 1D cross section or "line section", $\lambda$. The three superposed islands together span a fraction of the baseline, i.e. $\lambda_{\rm{islands}} / \lambda_{\rm{base}} = 3s$, with $s$ the average scaling factor of a single island. Since the three islands form the baseline for the next level, we conclude that the line section decreases exponentially with increasing level: 

\begin{equation}
\label{eq:scaling_factor}
    \lambda_n = \lambda_0 \left( 3s \right)^n. 
\end{equation}

\noindent Since the zoom-in level correlates with the number of encounters (each time we zoom in the baseline is increased by unity), we expect that the line section as a function of number of encounters decreases exponentially. We test this hypothesis by measuring the line section from the data as follows. 
We sort the simulations by increasing impact parameter. Each data point with index $i$ is assigned its own differential line section defined by $\Delta \lambda_i = \frac{1}{2}\left(p_{\rm{i+1}} - p_{\rm{i-1}}\right)$. Then, the cumulative line section for having $N_e = n$ close encounters is given by 

\begin{equation}
    \lambda\left( N_e = n \right) = \sum_i \Delta \lambda_i \delta_{n, N_{e,i}},
\end{equation}

\noindent with $\delta_{i,j}$ the Kronecker delta function, and $N_{e,i}$ the number of encounters for the data point with index $i$.  
The resultant profiles are given in Fig.~\ref{fig:lambda_vs_ne}. We confirm that up to $N_e = 7$ the data is consistent with an exponential decrease. For higher $N_e$, the fractal regions of the Jacobi spectra are unresolved, causing potential edge-effects, which result in the differential line section being overestimated. For the resolved part of the data however, both the black hole system and Sun-Earth-Moon system, as well as both positive and negative impact parameters, produce statistically consistent profiles:

\begin{equation}
    \log_{10}\,\lambda\left( N_e \right) = \alpha N_e + \beta, 
\end{equation}

\noindent with averaged values of $\alpha = -0.69 \pm 0.04$ and $\beta = 0.47 \pm 0.17$. We calculate the scaling factor by solving $10^{\alpha N_e} = \left( 3s \right)^{N_e}$, resulting in $s = 0.068 \pm 0.006$. In other words, about 80\% of the baseline is cut out, while the remaining 20\% is elevated to the next level. Furthermore, this 20\% is divided asymmetrically over the three islands, but on average, each individual island scales down by a factor $s$. A mathematically rigorous derivation of the fractal dimension, which takes into account the asymmetric islands, is beyond the scope of this paper. A rough estimate is obtained by considering that each new level has a factor $N=3$ more islands,  and where each individual island is scaled down by a factor $s$. This gives a fractal dimension of about

\begin{equation}
    D = \frac{\log{N}}{\log{\frac{1}{s}}} = \frac{\log{3}}{\log{\frac{1}{0.068}}} \approx 0.4.
\end{equation}

The line section profile for having a closest encounter of a certain separation, $\Delta r$, is also of interest. We measure this profile directly from the data in a similar way as we did for the number of encounters:  

\begin{equation}
\label{eq:dlambda}
    \lambda\left( r_p = \Delta r \right) = \sum_i \Delta \lambda_i \delta\left( r_{p,i} = \Delta r \right),
\end{equation}

\noindent where $r_p$ is the pericenter distance, $\Delta r$ the distance that is being evaluated, and in this case $\delta$ is a boolean function which is 1 if the statement between brackets is true or 0 if false. In practice, we define bins in $\log_{10}\Delta r$ of size $d\log_{10}\Delta r = 0.25$ and let the boolean function test whether a system falls within a certain bin. 
Another way to determine the line section profile is to exploit the fractal structure of the Jacobi spectrum. In the top panel of Fig.~\ref{fig:pro_vs_ret}, we observed two dips corresponding to head-on collisional trajectories. Each higher-order encounter adds self-similar, but scaled down copies of these dips. The line section profile for close encounters can then be thought of as being determined by the shape of these features at the base level (i.e. first encounter), multiplied by a constant to take into account the infinite number of scaled down copies at the higher levels. This can be expressed as a geometric series:

\begin{equation}
    \label{eq:series}
    \lambda_{\rm{\infty}}\left( \Delta r \right) = \lambda_1\left( \Delta r \right) \sum_{i=0}^{\infty} \left( 3s \right)^i,
\end{equation}

\noindent Here, $\lambda_1$ is the line section profile derived from the spectrum for the first encounter (e.g. top panel of Fig.~\ref{fig:pro_vs_ret}), calculated according to Eq.~\eqref{eq:dlambda}. The variable $s$ corresponds to the scaling factor derived earlier, while the factor 3 stems from the 3 islands in the spectra. The geometric series converges to $\lambda_1$ times a prefactor given by

\begin{equation}
    \lambda_{\rm{\infty}}\left( \Delta r \right) = \frac{1}{1 - 3s} \lambda_1\left( \Delta r \right) = 1.26 \lambda_1\left( \Delta r \right).
\end{equation}

\begin{figure}
\centering
\begin{tabular}{c}
\includegraphics[height=0.30\textwidth,width=0.48\textwidth]{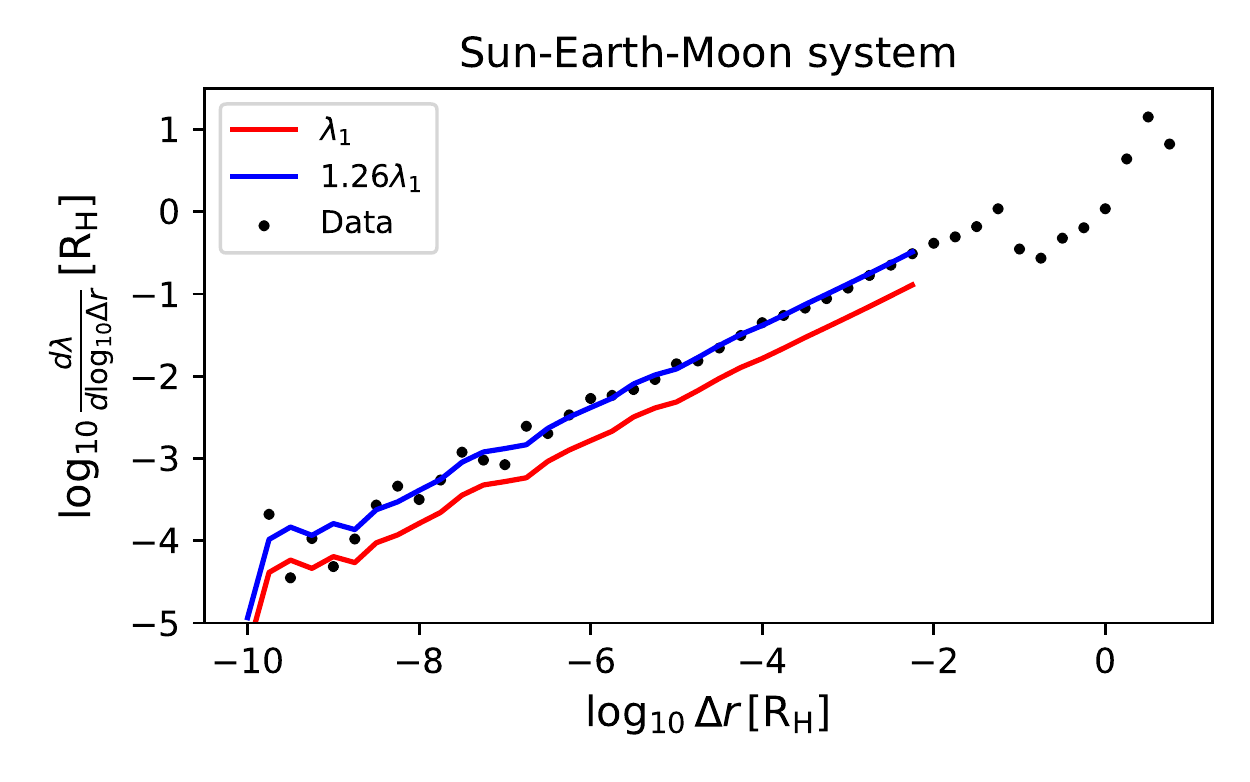} \\
\includegraphics[height=0.30\textwidth,width=0.48\textwidth]{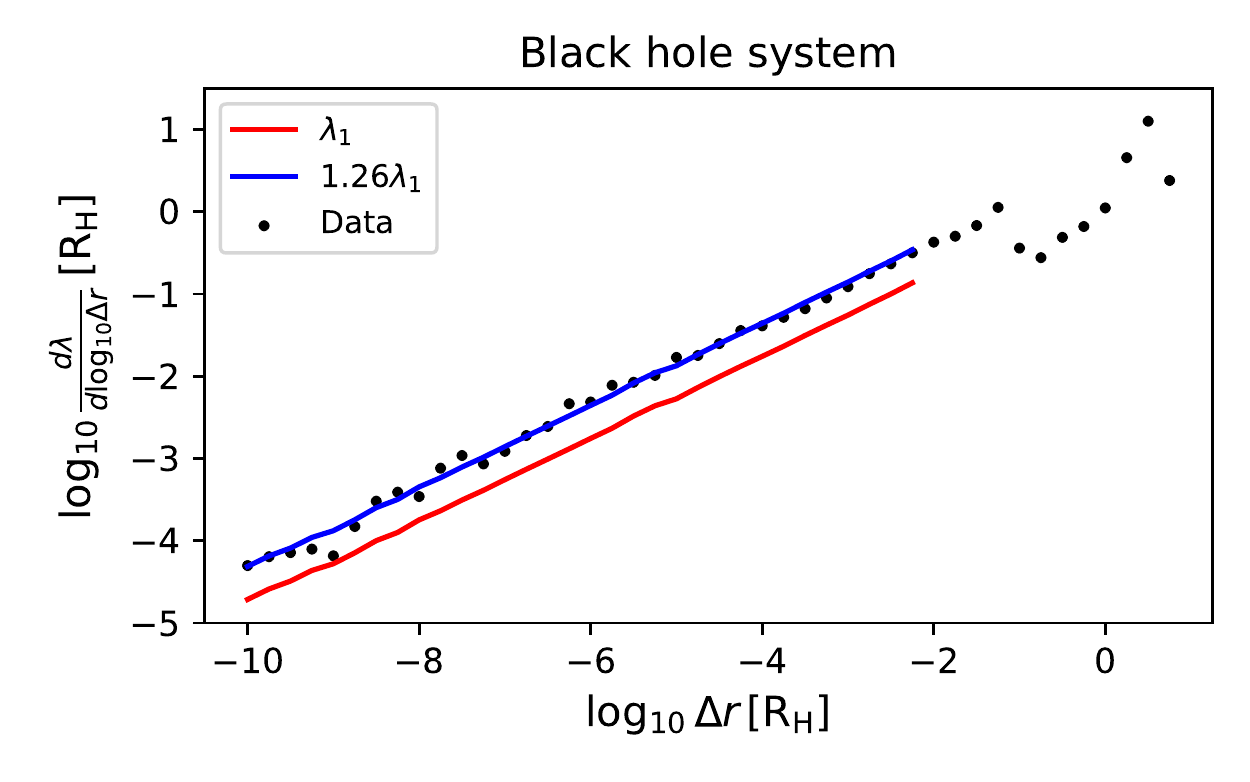} \\
\end{tabular}  
\caption{ Line section profiles for the closest separation, $\Delta r$, during the Jacobi capture. All profiles are consistent with a slope of 0.5, i.e. a square root growth. The different profiles are obtained with the full data set (black dots), first encounter spectrum (red line), and the converged geometric series (blue line).  }
\label{fig:lambda_vs_dr_sem}
\end{figure}

\noindent In other words, Jacobi captures increase the line section by about 26\% compared to single flybys only. In Fig.~\ref{fig:lambda_vs_dr_sem}, we plot the three line section profiles obtained by 1) Eq.~\eqref{eq:dlambda} for the closest encounter spectrum, 2) Eq.~\eqref{eq:dlambda} but only for the first encounter spectrum, i.e. $\lambda_1$, and 3) $\lambda_1$ multiplied by the prefactor of 1.26. We do this for both the Sun-Earth-Moon system and the AGN system. Below a hundredth of a Hill radius, the profiles are consistent with a square-root growth, i.e. the average of both fits to the full data set gives

\begin{equation}
    \label{eq:f_dr}
    \log_{10}\,\frac{d\lambda}{d\log_{10}\,\Delta r} = \mu \log_{10}\,\Delta r + \nu,
\end{equation}

\noindent with $\mu = 0.51 \pm 0.02$ and $\nu = 0.68 \pm 0.13$. 
By comparing the three different curves in each panel of Fig.~\ref{fig:lambda_vs_dr_sem}, we observe that the slope is indeed determined by the shape of the two characteristic dips in the Jacobi spectrum of the first encounter. 
An intuitive interpretation of the line section profile is obtained by regarding the close encounters as being approximately parabolic, where the closest separation $\Delta r$ is the periapsis distance $r_p$ 
(see also Sec.~\ref{sec:SEM}), with a uniform sampling of the angular momentum. The specific angular momentum of a parabolic orbit scales as $h \propto r_p^{{1}/{2}}$. The 
differential line section
then scales as $d\lambda = ({d\lambda}/{dh})\,({dh}/{dr_p})\, dr_p \propto r_p^{-{1}/{2}} dr_p$, implying that $d\lambda/d \log_{10} r_p = (\ln 10)r_p\,d\lambda/d r_p \propto r_p^{1/2}$, 
where we assumed a uniform sampling of $h$. 

\begin{figure}
\centering
\begin{tabular}{c}
\includegraphics[height=0.30\textwidth,width=0.48\textwidth]{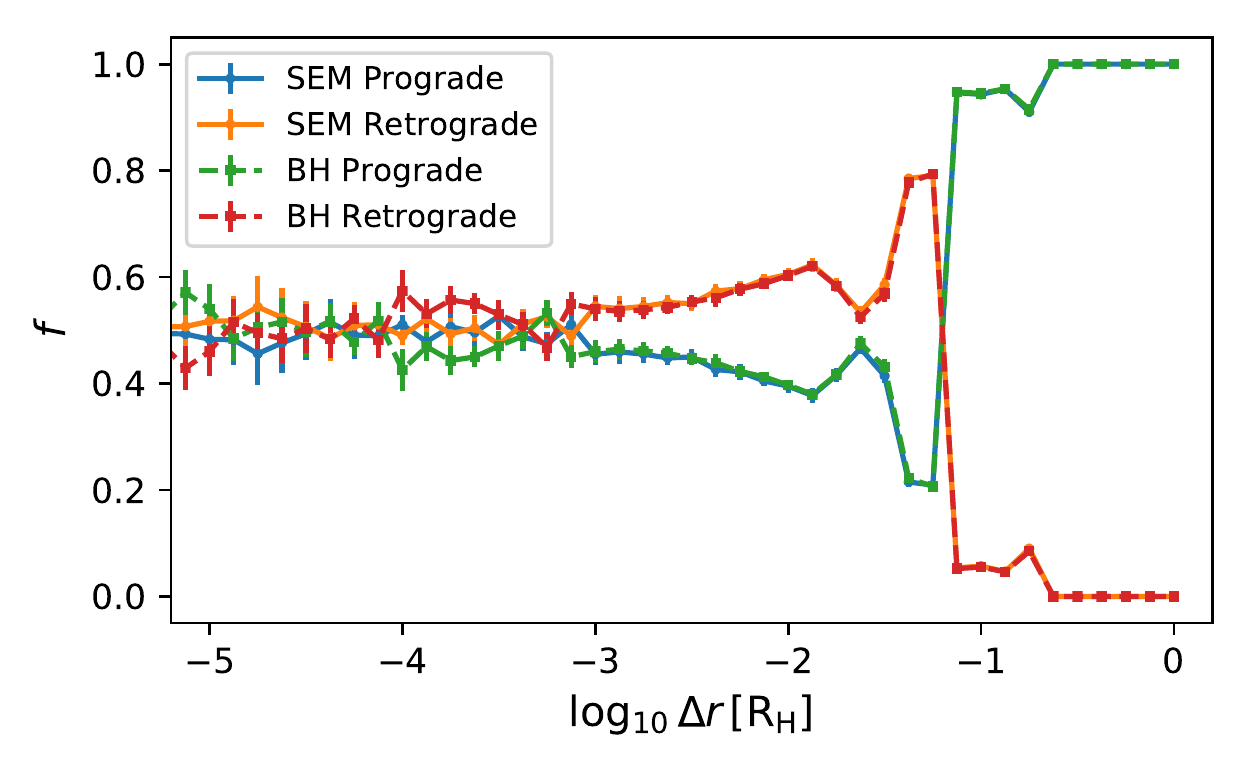} \\
\end{tabular}  
\caption{ Fractional contribution of prograde and retrograde encounters to the line section profiles given in Fig.~\ref{fig:lambda_vs_dr_sem}. }
\label{fig:lambda_vs_dr_gw}
\end{figure}

In Fig.~\ref{fig:lambda_vs_dr_gw}, we present the fractional contribution of prograde and retrograde encounters to the line section profiles in Fig.~\ref{fig:lambda_vs_dr_sem}. The errorbars correspond to the uncertainty in estimating the line section, i.e. in Eq.~\eqref{eq:dlambda} we replace $\Delta \lambda_i$ by the absolute value of a random value drawn from a Gaussian distribution with a standard deviation of $\Delta \lambda_i$. For this figure, we only include encounters within a Hill radius. However, for encounters further out, we observe in Fig.~\ref{fig:pro_vs_ret} that for impact parameters $p > 2.5$ all encounters are retrograde. This is expected as gravitational focusing is weak in this regime. For slightly closer encounters, the trajectories become of type $C$ in Fig.~\ref{fig:orbit_types}, and thus mostly prograde. 
Next, we observe an interesting transition from these wide and prograde encounters, to encounter separations of order $10^{-2}$ Hill radii being predominantly retrograde. 
For encounter separations smaller than about $10^{-4}$ Hill radii, the curves are rather noisy but flat, with an approximate equipartition between prograde and retrograde encounters. 
This last observation is interesting with respect to potential asymmetries between prograde and retrograde captures. If the dissipation mechanism depends steeply on the separation between the two bodies (e.g. for tidal dissipation and gravitational wave emission), it is most likely that the capture occurs during the closest encounter of the Jacobi capture. If the bodies are relatively large in units of their Hill radius (e.g. for the Earth-Moon), then there is a modest preference for retrograde captures. On the other hand, if the bodies are very small (e.g. for GW captures of black holes in AGN), then our data suggests an even partition between prograde and retrograde captures. 

So far, the characterisation of the closest encounters has mainly focused on the separation. The relative speed also plays a crucial role, which is encapsulated in the eccentricity distribution. Although the target and projectile bodies are not orbiting on isolated and closed Keplerian trajectories, we can nevertheless approximate this to be the case during very close approaches. We construct eccentricity distributions by collecting all Jacobi captures with $N_e = n$ encounters, evaluating the eccentricity of each encounter, and joining them into one ensemble. In Fig.~\ref{fig:ecc_histos}, we plot the results for $n=2,4,8$ and $16$. As expected, we find the encounters to be highly eccentric. The distribution is more consistent with a super-thermal eccentricity distribution than a thermal one.   

\subsection{Building a physical picture}\label{sec:physical_picture}

Based on our Figs.~\ref{fig:orbit_types} and \ref{fig:self_similar_ne}, we are able to contribute towards the development of a physical picture of Jacobi captures. 
Namely, one can distinguish 3 orbit families, which are visualised in Fig.~\ref{fig:orbit_types}, i.e. the red, green and blue families. They are different through their orientation during the closest approach (from the left, right or head-on in the figure). These 3 orbit-families correspond directly to the same-coloured islands in the dynamical spectra (Figs.~\ref{fig:spectra_ne}, \ref{fig:spectra_dr} and \ref{fig:self_similar_ne}). Furthermore, the first few zoom-in levels in Fig.~\ref{fig:self_similar_ne}, demonstrate that the same 3 islands are reproduced. This implies that a system can first belong to one orbit family, but can switch during the next encounter. 
As suggested by \citet{Petit86}, an extrapolation of this result to higher zoom-in levels would lead to the conclusion that there is a fractal, Cantor-like structure. 

For those outgoing orbits which manage to return to have another close encounter, it is thus possible to remain in the same orbit family (same color), or to switch to one of the other two families. The latter can be interpreted as the crossing of unstable periodic orbits to another domain in phase space. Each subsequent close encounter increases the order of the transition, as defined by  \citet{Petit86}. The self-similarity in our dynamical spectra is thus driven by the possibility of 1) having a next close encounter, and 2) switching between orbit families. This process continues until the final escape of the two bodies from their mutual Hill radius. The exponential lifetime distribution measured in Fig.~\ref{fig:lambda_vs_ne} can be interpreted by considering there is a fixed probability for escape after each close encounter during the Jacobi capture. Hence, it might be possible to model Jacobi captures statistically by using iterative maps, which take into account this escape probability, and a probability to switch between orbit families. Iterative maps are at the base of most famous fractal objects.

\section{Binary formation by dissipative Jacobi captures}\label{sec:dissipative}

\begin{figure}
\centering
\begin{tabular}{c}
\includegraphics[height=0.384\textwidth,width=0.48\textwidth]{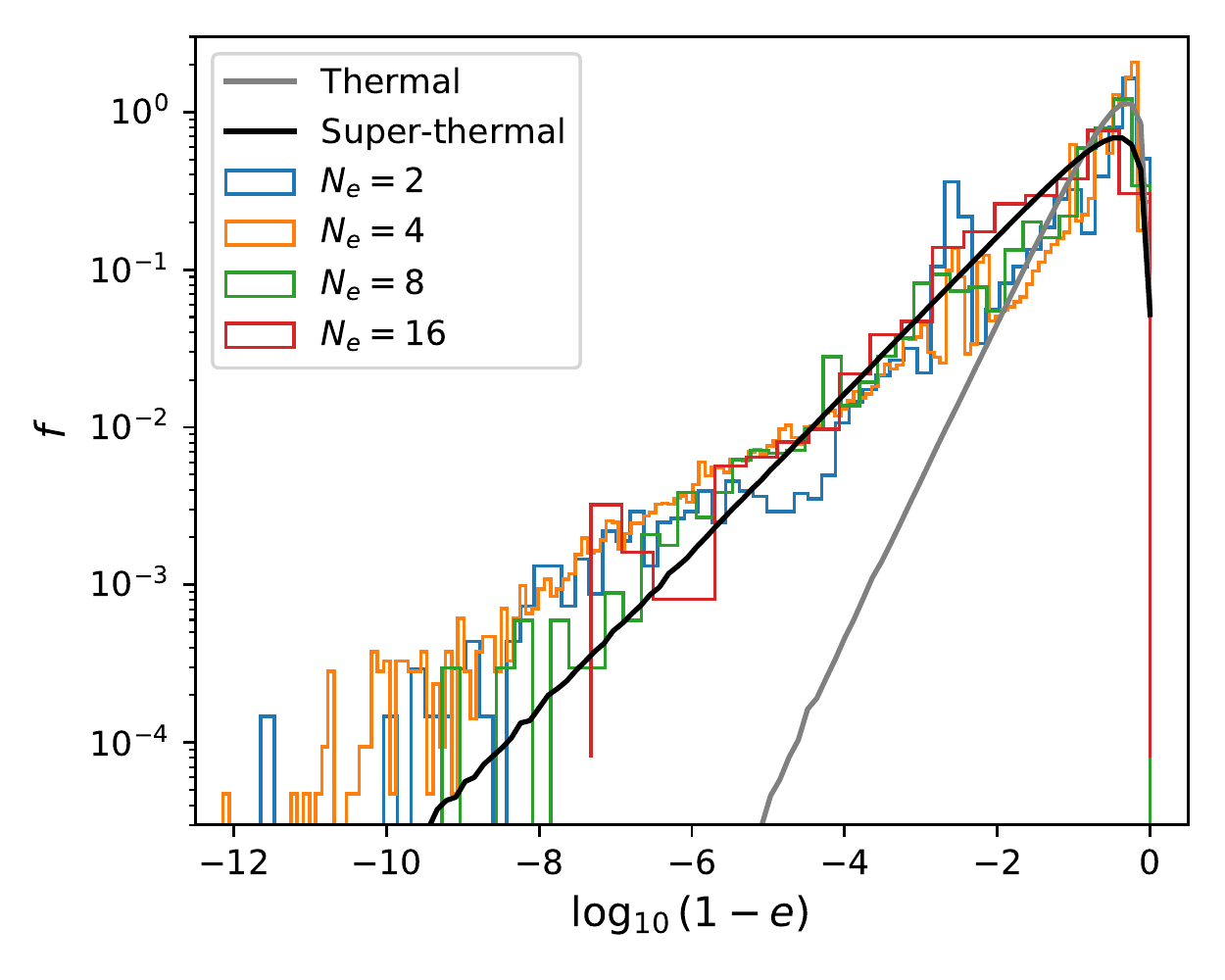} \\
\end{tabular}  
\caption{ Stacked eccentricity distributions for a given total number of encounters specified by $N_e$. As a benchmark, we also plot the thermal and super-thermal distributions.   }
\label{fig:ecc_histos}
\end{figure}

Having studied the conservative dynamics of Jacobi captures, we now focus on dissipative Jacobi captures as a channel for binary formation. We
post-process the effect of dissipation, rather than including it self-consistently during the numerical integration. In our first application, we regard a Jacobi capture of the Moon by the Earth and Sun, and consider both tidal capture and giant impacts. In our second application, we model the gravitational wave capture of stellar-mass black holes around a supermassive black hole. Realistic hydrodynamics simulations of gas embedded stellar-mass black holes in AGN discs, are discussed in our follow up study by \cite{Rowan_2021} (and references therein). 

\subsection{Jacobi captures with tidal dissipation: dynamical channel for giant impacts and tidal captures}\label{sec:SEM}

\begin{figure}
\centering
\begin{tabular}{c}
\includegraphics[height=0.24\textwidth,width=0.48\textwidth]{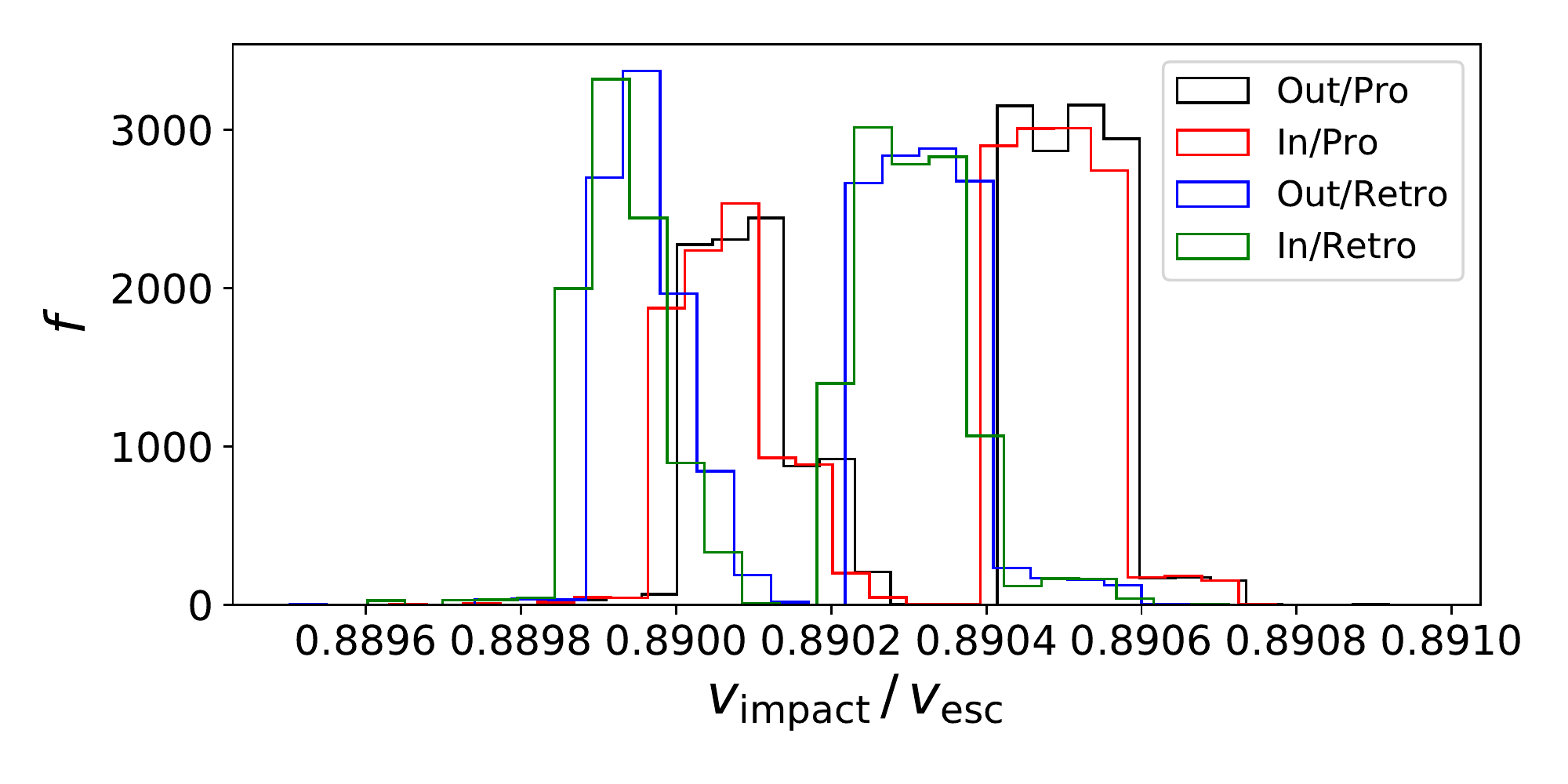} \\
\includegraphics[height=0.24\textwidth,width=0.48\textwidth]{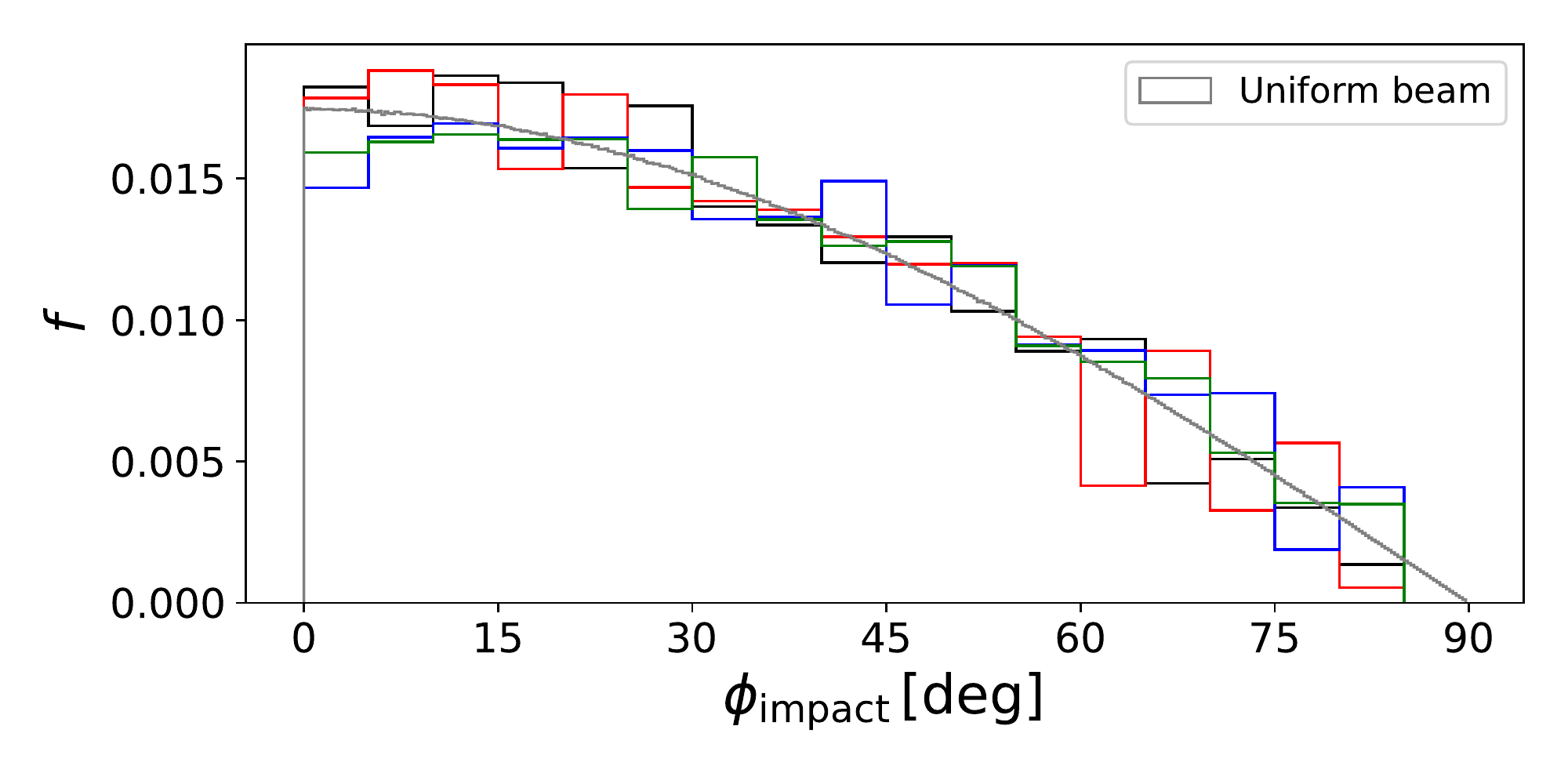} \\
\end{tabular}  
\caption{ Giant impact parameters for the Earth and the Moon during Jacobi captures. We plot the distributions of the impact speeds ($v_{\rm{impact}}$ normalised by Earth's escape speed $v_{\rm{esc}}$, top panel), and impact angles ($\phi_{\rm{impact}}$, where $0$ degrees corresponds to a head-on impact and $90$ degrees to barely grazing ones, bottom panel). We distinguish four ensembles depending on whether the Moon initially approached the Earth from the outside ("Out", i.e. further away from the Sun), or from the inside ("In"), and whether the final encounter that led to the impact was prograde ("Pro") or retrograde ("Retro"), with respect to Earth's orbit around the Sun.   }
\label{fig:impacts}
\end{figure}

From our numerical simulations of the Sun-Earth-Moon system with varying impact parameters, we extract those which led to a giant impact. Although the Moon-forming impactor, called Theia, is thought to have been a Mars-sized body, we will still adopt the mass and radius of the current-day Moon, in line with the tidal capture scenario. Hence, our calculated line sections for impacts 
are lower limits. 
For each of the giant impact events, we calculate the impact speed and impact angle at the moment the Earth and Moon first touch. The impact angle is defined as the angle between the relative velocity vector (in the frame where the Earth is at rest), and the relative position vector (pointing towards Earth). The angle ranges from 0 degrees for head-on collisions, to 90 degrees for barely grazing impacts. In Fig.~\ref{fig:impacts}, we plot the normalised histograms. The non-uniform sampling of the impact parameter space is corrected for by using the differential line sections as weights. To first order, we find that the impact speed ($9.95-9.97\,\rm{km\,s^{-1}}$) is given by the relative speed for a parabolic trajectory, at a distance of the sum of the radii of the Earth and Moon. A minor degeneracy is introduced by differentiating between prograde and retrograde encounters. The impact angles for the various types of approaches are mutually consistent. The distribution is identical to that of a uniform beam hitting a circular target. Constraints on the initial conditions for a Moon-forming impact, such as 1) impact speed of order Earth's escape speed, and 2) impact angle of about 45 degrees \citep{2004Icar..168..433C}, are thus naturally produced by Jacobi captures. 

Next, we consider the relative probability for giant impacts and tidal capture as a function of heliocentric distance. The line section for giant impacts is calculated analogous to Eq.~\eqref{eq:dlambda}, except that the boolean function now checks whether there was a close encounter separation smaller than the sum of their physical radii, i.e. a giant impact. To calculate the tidal capture line section, we require a formula for the tidal dissipation in highly eccentric binaries. However, the physics of tidal dissipation in rocky bodies is still uncertain, in particular in the high eccentricity regime. We adopt the relatively simple prescription derived by \cite{Wisdom_2008} for the mean energy dissipation rate per orbit due to tides, in binaries with arbitrary eccentricity (but zero obliquity):

\begin{equation}
    \label{eq:dedt_tidal}
    \frac{dE}{dt} = 21 \pi \frac{k_2}{Q} \frac{G M^2 R^5}{a^6 P} \zeta(e).
\end{equation}

\noindent Here, $k_2$ is the potential Love number of the tidally perturbed body, $Q$ is its effective tidal dissipation parameter and $R$ its radius. Furthermore, $M$ is the perturber's mass, $G$ the gravitational constant, $a$ is the semi-major axis of the orbit, $P$ is the orbital period, 
and $\zeta(e)$ is a function of the orbits' eccentricity \citep[see][Eq.~23 and 24]{Wisdom_2008}. In accordance with this model, we assume that the spins are synchronous with the orbit. 
This assumption leads to an upper limit as the rate of tidal dissipation for synchronous rotation is always greater than for asymptotic non-synchronous rotation, at zero obliquity \citep{Wisdom_2008}.
We ignore the tidal influence of the Sun on the Earth and the Moon. 
We adopt the following tidal parameters: $k_{2,Earth} = 0.302$ \citep{Wahr_1981}, $Q_{Earth}$ = 280 \citep{Ray_2001}, $k_{2,Moon} = 0.024059$ \citep{Konopliv_2013} and $Q_{Moon}$ = 37.5 \citep{Williams_2014}. These values were also tabelised by \cite{Lainey_2016}. We assume that the tidal parameters are constants, but there are studies indicating that they could have been different in the past \cite[e.g. see][]{Williams2000}.    
Both Earth and the Moon experience tides, and therefore both contribute to the total energy dissipation: 

\begin{equation}
    \label{eq:dedt_tot}
    \frac{dE}{dt} = \frac{dE}{dt}\Bigr|_{\rm{Earth}} + \frac{dE}{dt}\Bigr|_{\rm{Moon}}.     
\end{equation}

\noindent The ratio of Earth's dissipation to that of the Moon is 0.17, and therefore lunar tides dominate. 

For the simulations which did not lead to a giant impact, we calculate the Keplerian elements at the moment of closest approach, assuming that the target and projectile bodies are approximately isolated. If the eccentricity is larger than unity, or if the apocenter is larger than the Hill radius, then we make an approximation by setting the apocenter to be the Hill radius. In that case, the semi-major axis, $a$, and eccentricity, $e$, are recalculated from the pericenter and adjusted apocenter distances. 
Using the values of $a$ and $e$, and the tidal parameters given above, we are able to evaluate Eqs.~\eqref{eq:dedt_tidal} and \eqref{eq:dedt_tot}.
We calculate the post-encounter energy by subtracting the dissipated energy from the initial orbital energy, where $\Delta E = \frac{dE}{dt} P$. 
We will consider the Moon to be permanently captured if its new semi-major axis has decreased to below a fraction of the Hill radius: $a < f R_H$. The value of $f$ is varied from 0.5 to 0.25 in order to assess its sensitivity on the result.
The present day value is approximately $f=0.26$.

\begin{figure}
\centering
\begin{tabular}{c}
\includegraphics[height=0.567\textwidth,width=0.48\textwidth]{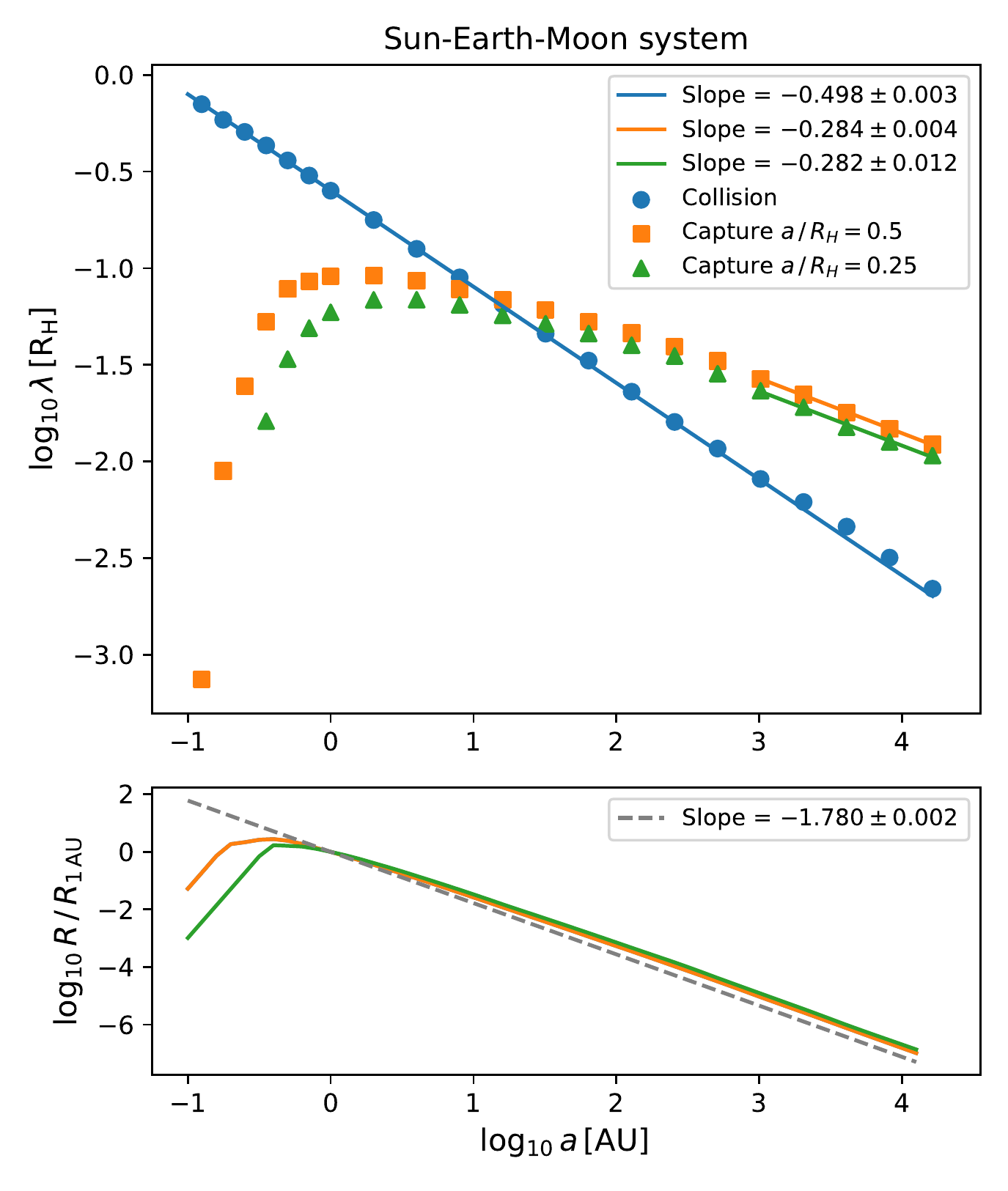} \\
\end{tabular}  
\caption{ Line section profile for collisions and tidal captures (top), and an estimate for the relative capture rate profile (bottom). The legend provides fits to the slopes of the data. The value of $a/R_H$ refers to the condition for permanent capture. The profiles exhibit a transition from collision dominated to tidal capture dominated dynamics.  }
\label{fig:lambda_vs_a_sem}
\end{figure}

The resulting line sections are presented in Fig.~\ref{fig:lambda_vs_a_sem}, which are normalised by the Hill radius. We observe that the line section for giant impacts decreases with heliocentric distance. While the Hill radius increases linearly with heliocentric distance, the physical radii of the bodies remain constant. At larger heliocentric distance, increasingly close encounters (in units of Hill radius) are required to still obtain giant impacts. The line section for impacts decreases as a power law with index $-\frac{1}{2}$. which follows from the line section profile for the closest separation (Fig.~\ref{fig:lambda_vs_dr_sem}). 
We note that in the three-dimensional case, the closest separation distribution is expected to be steeper (power law index of $-1$ instead of $-\frac{1}{2}$ \citep{Li2022}).

At heliocentric distances larger than $10$\,AU, the line section for tidal capture dominates. The slope in this region is approximately 
$\lambda_{\rm capture}/R_H \propto a^{-2/7}$.
At a heliocentric distance of order 10\,AU, there is a transition in the sense that at smaller distances collisions dominate, while further out tidal captures dominate the line section magnitudes. The tidal capture line section profile peaks near a heliocentric distance of 1\,AU. Below this value, there is a sharp decrease due to the high probability of collisions, while at larger radii, the line section decreases because the Hill radius increases faster than the tidal capture radius.

\begin{figure}
\centering
\begin{tabular}{c}
\includegraphics[height=0.567\textwidth,width=0.48\textwidth]{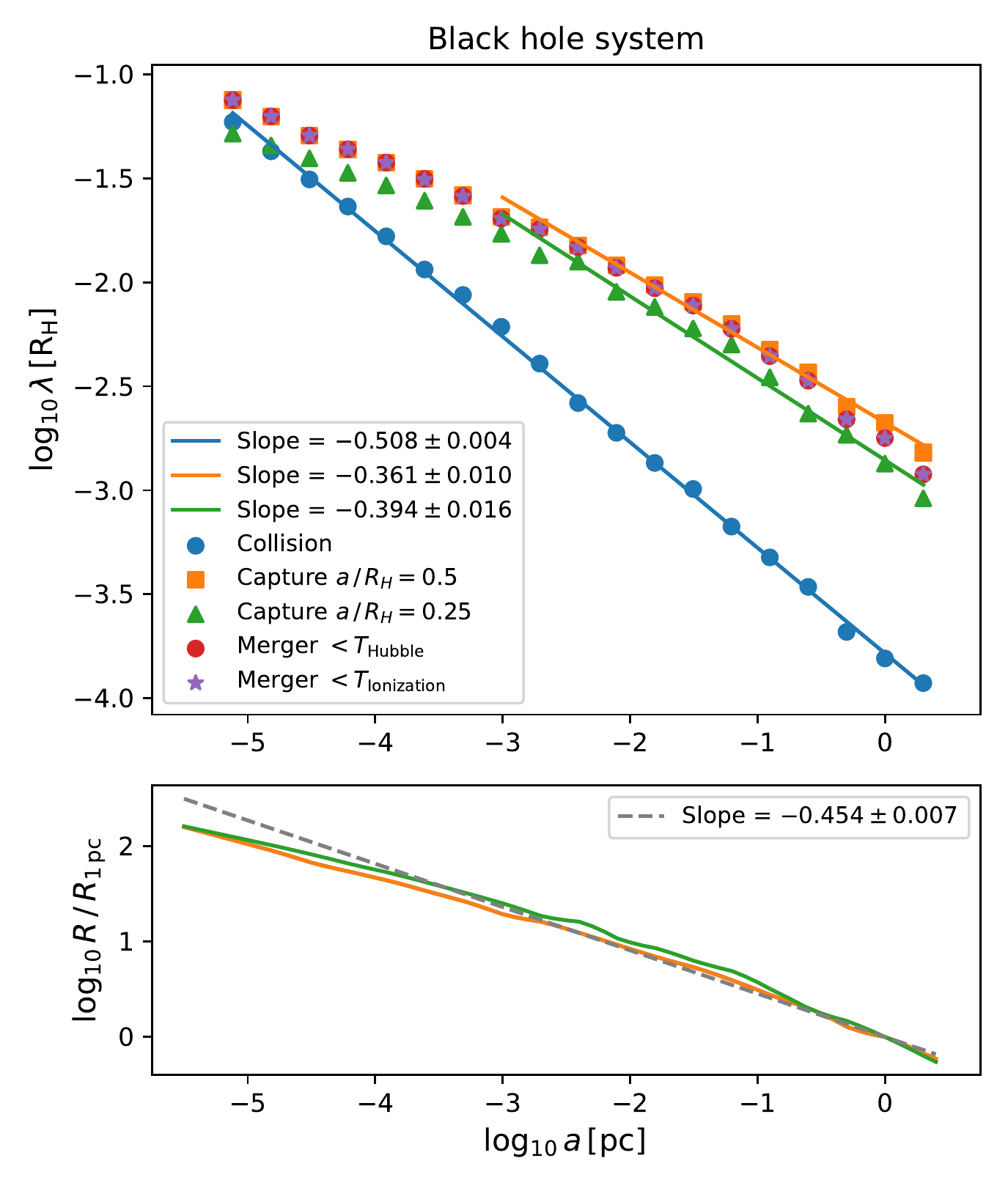} \\
\end{tabular}  
\caption{ Line section profile for collisions and gravitational wave captures for the black hole system (top), and an estimate for the relative capture rate profile (bottom). The legend provides fits to the slopes of the data. The value of $a/R_H$ refers to the condition for permanent capture. Gravitational wave sources are more efficiently formed in the inner parts of AGN discs through gravitational wave capture.} 
\label{fig:lambda_vs_a_gw}
\end{figure}

However, the probability of capture as a function of orbital radius in the disc, is best quantified through a relative rate, which is also dependent on the number density of candidate Moons and their relative velocity, which are both functions of heliocentric distance. As a toy model, we assume relative velocities given by the Keplerian shear ($v \propto a^{-\frac{1}{2}}$), and a local number density of 1 Moon per Hill radius ($n \propto a^{-2}$). This results in a rate which scales with heliocentric distance as $R\left(a\right) \propto a^{-\frac{25}{14}}$. 
The rate profiles associated with the numerical line sections and the analytical fit, are plotted in the bottom panel of Fig.~\ref{fig:lambda_vs_a_sem}. 
These profiles indicate that tidal captures are most likely to occur in the inner parts of planetary discs, but not too close to the host star, such that collisions dominate.
Concerning the giant impact of the Moon, we find a slight preference for a retrograde encounter. The sum of the radii of the Earth and Moon divided by the binary Hill radius is about $5 \times 10^{-3}$, and for this value we see in Fig.~\ref{fig:lambda_vs_dr_gw} that retrograde orbits are slightly more probable (about 60\% vs. 40\% for prograde encounters). Head-on and slightly retrograde impacts generate the most successful Moon-forming discs \citep{2012Sci...338.1047C}. 

\subsection{Jacobi captures with relativistic dissipation: formation channel for gravitational wave sources}\label{app:gw}

We perform a similar analysis as in the previous section, but instead apply it to the captures and mergers of black holes in AGN. The first difference is that the sizes of the black holes in units of their Hill radius is much smaller than for the Sun-Earth-Moon system, i.e. $10^{-9}-10^{-6}$ for orbital radii between $10^{-3}-1$ parsec, for our particular set of black hole masses (see Sec.~\ref{sec:methods}). 
The second difference is the source of dissipation, and therefore a different prescription for the amount of energy loss per orbital period. From \cite{Turner_1977}, we adopt the amount of dissipated energy per orbital period due to gravitational waves:

\begin{equation}
\label{eq:dE}
\Delta E = \frac{8}{15} \frac{G^{\frac{7}{2}}}{c^{5}} \left( m_t + m_p \right)^{\frac{1}{2}} \left(m_t m_p\right)^2 r_p^{-\frac{7}{2}} g\left( e \right), 
\end{equation} 

\noindent with $c$ the speed of light, $r_p$ the closest encounter separation, and $g\left(e\right)$ a function of eccentricity \citep[see Eqs. 29 and 31a in][]{Turner_1977}. Appropriate expressions for $g\left( e \right)$ are available for both bound and unbound orbits. Contrary to Eq.~\eqref{eq:dedt_tidal}, this expression includes an explicit dependence on the pericenter distance, which we extract directly from the simulations. The eccentricity is calculated in the isolated, two-body approximation from the pericenter distance and speed during the closest approach. Similar to what we did for the Sun-Earth-Moon experiment, we calculate the post-encounter orbital energy and determine if the binary is bound, and if its new semi-major axis is less than a fraction, $f$, of the Hill radius. If these constraints are met, we consider the capture to be a permanent one, and its differential line section is added to the cumulative line section for gravitational wave capture. For each post-encounter binary we also calculate the time to merger according to \cite{Peters_1963}. We construct two extra line section profiles by adding the constraint that 1) the binary has to merge within a Hubble time, and 2) the binary has to merge within the ionisation time scale, i.e. the time to breakup due to an interaction with another body and which is approximated to be 100 Myr \citep{Bartos17}. 
In App.~\ref{sec:appendix}, we construct a toy model for an AGN disc with the aim of determining where in the disc dissipative Jacobi captures are most efficient.
The resultant numerical line section and relative rate profiles are given in Fig.~\ref{fig:lambda_vs_a_gw}. 

Due to the fact that black holes are much more compact in units of their Hill radius compared to the Earth-Moon system, we find that the line section is dominated by gravitational wave capture as opposed to direct collisions. 
The slope of the line section profiles for capture is not a constant, but a fit to the data between radii of $10^{-3}$ to 1 parsec produces slopes of approximately $-5/14$. We derive this value analytically in Appendix~\ref{sec:appendix}. Furthermore, we observe that the profiles with the added merger time constraints barely differ from the profile without time constraints. This can be understood by considering the extremely high eccentricities of the captured binaries (Fig.~\ref{fig:ecc_histos}), which reduce the gravitational wave in-spiral time scale.

\section{Conclusions and caveats}\label{sec:caveats}

In our first systematic study on the dynamics of Jacobi captures, we introduced several simplifying assumptions. We reduced the geometry to be confined to two dimensions, and we assumed initially circular orbits for the target and projectile bodies. Such a model already reveals a rich dynamics. The phase space structure is self-similar and can be described by a generalised Cantor set with a dimension of approximately 0.4 (see Sec.~\ref{sec:3.3}). The lifetime of Jacobi captures is described by an exponential decay (see Fig.~\ref{fig:lambda_vs_ne}), while the line section for close approaches between the target and projectile bodies follows a power law distribution with index 0.5 (see Fig.~\ref{fig:lambda_vs_dr_sem}). The importance of Jacobi captures relative to single flybys is quantified by their relative contribution to the line sections: about 20\% for Jacobi captures vs. 80\% for single flybys (see Sec.~\ref{sec:3.3}). 

Asymmetries between approaches from the inner or outer disc are negligible (see Figs.~\ref{fig:spectra_ne}, \ref{fig:spectra_dr} and \ref{fig:impacts}). There is however an interesting asymmetry between prograde and retrograde orientations for relatively wide encounters (see Fig.~\ref{fig:pro_vs_ret} and \ref{fig:lambda_vs_dr_gw}). For separations of $10^{-1} < \Delta r\,/\,R_H < 1$, encounters are predominantly prograde. For $10^{-4} < \Delta r\,/\,R_H < 10^{-1}$, we find a slight preference for retrograde enccounters (60\% retrograde vs. 40\% prograde). For closer separations however, there is an equipartition between prograde and retrograde encounters. 

Applying these results to the Earth-Moon system, we find a slight preference for a retrograde pre-impact encounter. Furthermore, Jacobi captures produce impact speeds of order Earth's escape speed, while the median impact angle is 30 degrees, with the 10\% and 90\% intervals given by 6 and 64 degrees (see Fig.~\ref{fig:impacts}). 
These results are an advantage for the Jacobi capture origin scenario, since head-on and slightly retrograde impacts generate the most successful Moon-forming discs \citep{2012Sci...338.1047C}. 
Furthermore, there is a transition region in heliocentric distance from 1-10\,AU, below which impacts are most likely to occur, while beyond, tidal captures become the dominant outcome (see Fig.~\ref{fig:lambda_vs_a_sem}).
Jacobi captures with tidal dissipation thus form a promising scenario for explaining irregular moons around giant planets.  

Going beyond the Solar System, we find that Jacobi capture interactions between black holes in AGN can lead to the formation of binary black holes. Thereby, Jacobi captures have the potential to contribute to the AGN channel for producing gravitational wave sources. Right after the capture phase, we find that the binaries can be both prograde and retrograde {with respect to the AGN disc}, with an approximate super-thermal eccentricity distribution (see Fig.~\ref{fig:ecc_histos}). 
The very high eccentricities drive the rapid inspiral and circularisation due to gravitational wave emission.
The line section for permanent capture scales with the total mass as $\lambda \propto \left( m_p + m_t \right)^{\frac{2}{7}}$ and with the product of the masses as $\lambda \propto \left( m_t m_p \right)^{\frac{1}{7}}$ (see App.~\ref{sec:appendix}), thus slightly favouring massive mergers with similar mass components.
Our results motivate follow up studies of the AGN channel, in which some of our assumptions should be relaxed.

In more realistic discs than the ones considered here, bodies have a small but non-zero eccentricity and inclination distribution. This affects not only the approach angle between the two bodies, but also their relative speed. If the velocity dispersion of the disc dominates over the Kepler shear, we expect the efficiency of Jacobi captures to reduce, as the projectile body would simply fly through the Hill radius as if they were isolated. 
However, if the number density of bodies in the disc is large enough, and the relative velocity distribution exhibits a low-velocity tail, then a subset of bodies can still be captured.

Another caveat in our model is the post-processing of the dissipative forces. More realistic results can be obtained by implementing dissipation (tides, gravitational waves, etc.) directly into the numerical integration. Such simulations would also directly test our assumption that a binary is permanently captured if its post-encounter semi-major axis is a small fraction of the Hill radius, or whether a permanent capture is a more complicated process. An example of such a study is presented in an accompanying paper by \cite{Rowan_2021}, 
who take into account the gravitational influence and accretion of gas using the code \texttt{Phantom} \citep{2018PASA...35...31P}. Nevertheless, our post-processing method of incorporating tides and gravitational wave emission has shown that dissipative Jacobi captures have a high potential for producing a large variety of binary systems and enhancing mergers and collisions. 

We applied the mechanism of dissipative Jacobi captures to two different systems, the Sun-Earth-Moon system and black holes in galactic nuclei. Due to the universality of Newton's law of gravity and the ubiquity of disc configurations in the Universe, we expect Jacobi captures to occur in a wide variety of astrophysical systems, and on a large range of scales. Candidate configurations include planet-planet encounters in planetesimal discs, cometary flybys through planetary systems in the background of a galactic potential, and stellar encounters with giant molecular clouds in galactic discs. The dissipation source varies per astrophysical system, but has a direct influence on the outcome of Jacobi captures. For example, the physics of tides in stars and planets is still uncertain. By modelling the evolution and products of Jacobi captures, we can compare and scrutinise different dissipation models.     
We conclude that dissipative Jacobi captures provide a dynamical framework for explaining the origin of a wide variety of binary systems. Our results motivate follow up research into the rates, universality and other implications of Jacobi captures in astrophysical systems over a large range of scales.  \\   


\section*{Acknowledgements}

We thank Alexandre Correia for discussions on Jacobi captures in the Solar System, and Maria K. Peto for discussions on the geochemical arguments related to the formation of the Moon.  
We also thank the referee for suggesting improvements to the paper.
This project was supported by funds from the European Research Council (ERC) under the European Union’s Horizon 2020 research and innovation program under grant agreement No 638435 (GalNUC). 
  We used the following software: Brutus \citep{2015ComAC...2....2B},  
          AMUSE \citep{amuse18} and
          Python \citep{python95}.

\section*{Data Availability} 

The data underlying this article will be shared on reasonable request to the corresponding author.

\bibliography{agn}{}
\bibliographystyle{aasjournal}


\appendix

\section{Jacobi captures with gravitational wave emission}
\label{sec:appendix}

We consider a system consisting of two stellar mass black holes in orbit around a supermassive black hole (SMBH). The orbital energy, $E$, of the two black holes during a close approach within their Hill radius is approximated as

\begin{equation}
\label{eq:E}
E \approx -\frac{1}{2}\frac{G m_t m_p}{R_H}, 
\end{equation}

\noindent with $G$ the gravitational constant, $m_t$ and $m_p$ the mass of the target and projectile body respectively, and $R_H$ the Hill radius (Eq.~\eqref{eq:binRH}).
Substituting the expression for the Hill radius into Eq.~\eqref{eq:E}, we get

\begin{equation}
\label{eq:ErH}
E \approx -\frac{1}{2} \left( \frac{1}{3} \right)^{-\frac{1}{3}} G M^{\frac{1}{3}} \left( m_t + m_p \right)^{-\frac{1}{3}}  m_t m_p r^{-1},
\end{equation}

\noindent with $r$ the orbital radius from the SMBH. From \cite{Turner_1977}, we adopt the amount of dissipated energy per orbital period due to gravitational waves (see Eq.~\eqref{eq:dE}). 
The absolute fractional energy change is the ratio of Eqs.~\eqref{eq:dE} and \eqref{eq:ErH}, 

\begin{equation}
\label{eq:dE_E}
\left| \frac{\Delta E}{E} \right| \equiv f_E \approx \frac{16}{15} \left(\frac{1}{3}\right)^{\frac{1}{3}} \frac{G^{\frac{5}{2}}}{c^{5}} g\left( e \right) M^{-\frac{1}{3}} \left( m_t + m_p \right)^{\frac{5}{6}} m_t m_p r_p^{-\frac{7}{2}} r.
\end{equation}

\noindent For a gravitational wave capture to occur, the two black holes need to dissipate an amount of energy, such that the new semi-major axis is (much) smaller than the Hill radius. Given this amount of $\Delta E_c$, we define the associated separation as the gravitational wave capture radius, i.e. $r_p$ in Eqs.~\eqref{eq:dE} and \eqref{eq:dE_E} is replaced by the capture radius,  $r_{\rm{c}}$. We obtain an expression for the gravitational wave capture radius by rewriting Eq.~\eqref{eq:dE_E} for $r_{\rm{c}}$:

\begin{equation}
\label{eq:rpc}
r_c \approx C_1 M^{-\frac{2}{21}} \left( m_t + m_p \right)^{\frac{5}{21}} \left( m_t m_p \right)^{\frac{2}{7}} r^{\frac{2}{7}},
\end{equation}
 
\noindent where we gathered all constants (including $f_E$) and $g\left( e \right)$ into the prefactor $C_1$. In units of the Hill radius the capture radius is given by

\begin{equation}
\label{eq:rc_hill}
\frac{r_c}{R_H} \approx C_2 M^{\frac{5}{21}} \left( m_t + m_p \right)^{-\frac{2}{21}} \left( m_t m_p \right)^{\frac{2}{7}} r^{-\frac{5}{7}}, 
\end{equation}

\noindent with $C_2$ the new prefactor. 

Next, we require the cumulative line section for having an encounter closer or equal to the capture radius. We derive the cumulative line section profile by integrating Eq.~\eqref{eq:f_dr} from $\Delta r = 0$ to $\Delta r = r_c$, resulting in

\begin{equation}
\label{eq:lambda_cum}
\log_{10} \frac{\lambda_c}{R_H} = \mu \log_{10} \frac{r_c}{R_H} + \nu - \log_{10}\mu.
\end{equation}

\noindent Substituting in Eq.~\eqref{eq:rc_hill}, we get 

\begin{equation}
\label{eq:lc1}
\begin{split}
\log_{10} \frac{\lambda_c}{R_H} = \mu \left( \log_{10}\,C_2 + \frac{5}{21}\log_{10}\,M - \frac{2}{21}\log_{10}\,\left( m_t + m_p \right) \right.\\
\left.+ \frac{2}{7}\log_{10}\,\left( m_t m_p \right) - \frac{5}{7}\,\log_{10}\,r \right) + \nu - \log_{10}\mu.
\end{split}
\end{equation}

\noindent The predicted power law index for the dependence of $\lambda_c/R_H$ on $r$ is thus $-5\mu/7 \approx -5/14 \approx -0.357$. This value is consistent with the profiles for gravitational wave capture in Fig.~\ref{fig:lambda_vs_a_gw}. 
From this same figure, we use the data point at $a=1\,\rm{pc}$, for the profile with the mildest capture constraint, $a/R_H = 0.5$, to evaluate the constants in Eq.~\eqref{eq:lc1}. Furthermore, adopting the black hole masses from our experiments, we can rewrite Eq.~\eqref{eq:lc1} as

\begin{equation}
\begin{split}
\frac{\lambda_c}{R_H} = 2.1 \times 10^{-3} \left(\frac{M}{4\,\times\,10^6\,\rm{[M_\odot]}}\right)^\frac{5}{42} \left(\frac{m_t+m_p}{83\,\rm{[M_\odot]} + 65\,\rm{[M_\odot]}}\right)^{-\frac{1}{21}}\\
\left( \frac{m_t}{83\,\rm{[M_\odot]}} \frac{m_p}{65\,\rm{[M_\odot]}} \right)^\frac{1}{7} \left(\frac{r}{\rm{pc}}\right)^{-\frac{5}{14}},
\end{split}
\end{equation}

\noindent Since the Hill radius can be expressed as 

\begin{equation}
\begin{split}
R_H = 0.023\,\rm{[pc]} \left(\frac{M}{4\,\times\,10^6\,\rm{[M_\odot]}}\right)^{-\frac{1}{3}} \left(\frac{m_t+m_p}{83\,\rm{[M_\odot]} + 65\,\rm{[M_\odot]}}\right)^{\frac{1}{3}} 
\left(\frac{r}{\rm{pc}}\right),
\end{split}
\end{equation}

\noindent we can write the line section for capture as

\begin{equation}
\label{eq:l(r)}
\begin{split} 
\lambda_c\left( r \right) \approx 5 \times 10^{-5}\,\rm{[pc]} \left(\frac{M}{4\,\times\,10^6\,\rm{[M_\odot]}}\right)^{-\frac{3}{14}} \left(\frac{m_t+m_p}{83\,\rm{[M_\odot]} + 65\,\rm{[M_\odot]}}\right)^{\frac{2}{7}}\\
\left( \frac{m_t}{83\,\rm{[M_\odot]}} \frac{m_p}{65\,\rm{[M_\odot]}} \right)^\frac{1}{7} \left(\frac{r}{\rm{pc}}\right)^{\frac{9}{14}}.
\end{split}
\end{equation}

\noindent Using this expression for the capture line section, we are able to calculate an ``$n \lambda v$'' rate of gravitational wave-assisted Jacobi captures.
We require a surface number density profile for the black holes and a relative velocity profile. We assume that the projectile masses are distributed in the disc according to a power law profile:

\begin{equation}
\label{eq:n(r)}
n\left( r \right) = 223\,\rm{[pc^{-2}]} \left(\frac{2-p}{2-0.6}\right) \left(\frac{N_{p}}{10^3}\right) \left(\frac{R_{\rm{out}}}{\rm{pc}}\right)^{p-2} \left(\frac{r}{\rm{pc}}\right)^{-p}, 
\end{equation}

\noindent with $N_p$ the number of projectile bodies, $R_{\rm{out}}$ the outer radius of the disc, and $p$ the power law index. The nominal disc model will consist of $N_p = 10^3$, $R_{\rm{out}} = 1\,\rm{pc}$, and $p=0.6$.    
The relative velocity between the target and projectile bodies is estimated from the local circular velocity shear within the binary Hill radius:

\begin{equation}
\label{eq:v(r)}
\begin{split}
v\left( r \right) = 1.5\,\rm{[km\,s^{-1}]} \left(\frac{M}{4\times10^6\,\rm{[M_\odot]}}\right)^{\frac{1}{6}} \left(\frac{m_t+m_p}{83\,\rm{[M_\odot]} + 65\,\rm{[M_\odot]}}\right)^{\frac{1}{3}}
\left(\frac{r}{\rm{pc}}\right)^{-\frac{1}{2}}.
\end{split}
\end{equation}   

\noindent By combining Eqs.~\eqref{eq:l(r)}, \eqref{eq:n(r)} and \eqref{eq:v(r)}, we are able to estimate the rate of GW captures at a particular radius in the disc for a single target black hole:

\begin{equation}
\begin{split}
R\left( r \right) = n \lambda v = 1.7 \times 10^{-8}\,\rm{[yr^{-1}]} \left(\frac{2-p}{2-0.6}\right)
\left(\frac{M}{4\,\times\,10^6\,\rm{[M_\odot]}}\right)^{-\frac{1}{21}}\\
\left(\frac{R_{\rm{out}}}{\rm{pc}}\right)^{p-2}
\left(\frac{N_{p}}{10^3}\right) \left(\frac{m_t+m_p}{83\,\rm{[M_\odot]} + 65\,\rm{[M_\odot]}}\right)^{\frac{13}{21}}\left( \frac{m_t}{83\,\rm{[M_\odot]}} \frac{m_p}{65\,\rm{[M_\odot]}} \right)^\frac{1}{7}\\
\left(\frac{r}{\rm{pc}}\right)^{\frac{1}{7}-p}.
\end{split}
\end{equation}

\noindent We find that the dependence of the rate, $R$, on the orbital radius, $r$, is a power law with index $1/7-p$. For our nominal value of $p=0.6$ this gives an index of about $-0.457$. This value is consistent with the profile in the bottom panel of Fig.~\ref{fig:lambda_vs_a_gw}, which is obtained directly from the numerical line section profile. 


\end{document}